\documentclass[acmlarge]{acmart}
\AtBeginDocument{%
  \providecommand\BibTeX{{%
    \normalfont B\kern-0.5em{\scshape i\kern-0.25em b}\kern-0.8em\TeX}}}
\usepackage[utf8]{inputenc}
\usepackage{multirow}
\usepackage[table]{colortbl}
\usepackage{xcolor}
\usepackage{color,soul}
\usepackage{makecell}

\newcommand\crule[3][black]{\textcolor{#1}{\rule{#2}{#3}}}
\setcopyright{acmcopyright}
\copyrightyear{2023}
\acmYear{2023}

\acmJournal{CSUR}

\begin{document}

\title[ML Systems: A Survey from a Data-Oriented Perspective]{Machine Learning Systems:\\ A Survey from a Data-Oriented Perspective}
\author{Christian Cabrera}
\author{Andrei Paleyes}
\author{Pierre Thodoroff}
\author{Neil D. Lawrence}
\email{{chc79, ap2169, pt440, ndl21}@cam.ac.uk}
\affiliation{%
  \institution{\\Department of Computer Science and Technology, University of Cambridge}
  \city{Cambridge}
  \country{United Kingdom}
}

\renewcommand{\hl}[1]{{\color{black} #1}}
\newcommand{\hlr}[1]{{\color{black} #1}}
\newcommand{\hlg}[1]{{\color{black} #1}}
\newcommand{\hlb}[1]{{\color{black} #1}}

\begin{abstract}
Engineers are deploying ML models as parts of real-world systems with the upsurge of AI technologies. Real-world environments challenge the deployment of such systems because these environments produce large amounts of heterogeneous data, and users require increasingly efficient responses. These requirements push prevalent software architectures to the limit when deploying ML-based systems. Data-oriented Architecture (DOA) is an emerging style that equips systems better for integrating ML models. Even though papers on deployed ML systems do not mention DOA, their authors made design decisions that implicitly follow DOA. Implicit decisions create a knowledge gap, limiting the practitioners' ability to implement ML-based systems. \hlb{This paper surveys why, how, and to what extent practitioners have adopted DOA to implement and deploy ML-based systems.} We overcome the knowledge gap by answering these questions and explicitly showing the design decisions and practices behind these systems. The survey follows a well-known systematic and semi-automated methodology for reviewing papers in software engineering. The majority of reviewed works partially adopt DOA. Such an adoption enables systems to address requirements such as Big Data management, low latency processing, resource management, security and privacy. Based on these findings, we formulate practical advice to facilitate the deployment of ML-based systems.
\end{abstract}



\maketitle

\section{Introduction}
\label{sec:introduction}

Artificial Intelligence solutions based on ML have gained an increased level of attention. They have been used to address challenging problems in several domains (e.g., healthcare, science, etc.)~\cite{lecun2015deep,van2023ai}. Their success is a product of the growth of available data, increasingly powerful hardware, and the development of novel ML models~\cite{dovsilovic2018explainable,raiaan2024surveyllms}. \hlr{Many of these models have demonstrated practical value, which has led to their rapid adoption in real-world software systems. The contrast between real-world environments and the more controlled environments from which ML models originate challenges deploying systems that integrate these ML models~\cite{cabrera2024s4}}. In particular, real-world environments produce large amounts of complex, dynamic, and sometimes sensitive data, which systems must process efficiently~\cite{joshi2007data,cabrera2022self}. ML-based systems deployed in the real world must be scalable, autonomous, and resource-efficient, while enabling data availability, monitoring, security, and trust~\cite{polyzotis2018data, paleyes2022challenges, Lwakatare2020LargescaleML}. 

Service-oriented architectures (SOAs) and microservices are the dominant \hlg{software architectural styles} nowadays~\cite{oreilly2020microservices, aniche2019current}. \hlg{These styles are characterised} by services that separate systems' functionalities, communicate through well-defined programming interfaces (i.e., APIs), and are usually deployed in the cloud~\cite{microservices2018patterns}. The necessity to process large volumes of data, typical for ML-based real-world systems, leads to new requirements that challenge SOA-based systems. For example, separation of concerns facilitates systems maintenance by following the divide-and-conquer principle~\cite{microservices2016dividends, microservices2019challenges}. However, data monitoring and transparency requirements are arduous to satisfy because services hide the systems' data behind their APIs~\cite{cobbe2023accountability}. This situation is known as ``The Data Dichotomy'': while high-quality data management requires exposing systems' data, services hide it behind their interfaces~\cite{stopford2016data}. Another example is that services deployed in flexible cloud platforms support highly available and scalable systems~\cite{microservices2021deployment}. However, cloud deployments struggle to meet low-latency requirements for critical applications. The physical distance between end users and cloud data centres impacts systems' end-to-end response time~\cite{cabrera2023maaco}. Such cloud-based deployments also affect data security, privacy, and trust requirements because the ownership of the data changes from end users to cloud providers. This change of ownership also generates availability, privacy and security concerns that can impact ML model training~\cite{Kleppmann2019localfirst,akbar2023security,cabrera2024syseng}.

Data-Oriented Architecture (DOA) is an emerging \hlg{architectural style} that complements the current practices for implementing and deploying systems. DOA considers data as the common denominator between system components~\cite{joshi2007data}. Services in DOA are distributed, autonomous, and communicate with each other at the data level (i.e., data coupling) using asynchronous messages~\cite{schuler2015data,vorhemus2017data}. DOA enables systems to achieve data availability, reusability, monitoring, and systems autonomy and resource efficiency~\cite{joshi2007data,schuler2015data,vorhemus2017data}. These properties facilitate systems integrating ML models in production environments. For example, the data coupling between DOA-based system components enables the storage of the states of the data that flows through it by design. A practitioner or a meta-learning system can analyse such records, identify when the system's data distribution changes, and update the system's learning models. Similarly, DOA enables the creation of open systems because of the autonomous communication between its components. New devices can automatically join and provide computing power to DOA-based systems. Such flexibility facilitates the deployment of ML models in \hlr{resource-constrained} environments and can mitigate low-latency requirements as services are physically closer to end users~\cite{cabrera2021rl,tabatabaee2022mecsurvey}.

\hlb{Even though most papers on deployed ML-based systems do not describe DOA explicitly, their authors faced the same challenges that DOA aims to address. Such an implicit adoption creates a knowledge gap that limits the practitioners' ability to deploy ML-based systems in real-world conditions. We close this gap in this survey paper by explicitly showing the design decisions behind implementing and deploying ML-based systems and analysing why, how and to what extent their authors have implicitly adopted DOA principles. Such an analysis constitutes the first in-depth study of DOA architectural style for ML-based systems.} \hlr{Based on existing literature on DOA, we formulate high-level principles of this architectural style and contextualise these principles through the ML deployment challenges. We then conduct a comprehensive survey to understand the implicit adoption of DOA principles among practitioners when implementing existing ML-based systems\footnote{\hlr{We refer to "ML-based systems in the real world" as software systems that deliver ML-powered capabilities deployed and evaluated in production environments, with real users, unpredictable physical processes, unseen variable data. Such a definition contrasts against isolated academic settings with fixed datasets, simulated users and environments, and tightly controlled experimental conditions.}}. Based on the survey findings, this paper shows the benefits and limitations of DOA, formulates practical advice for deploying ML-based systems, and discusses open challenges and research directions to advance DOA.}
\section{Related Work}
\label{sec:related-work}

Although DOA as an architectural style for ML-based systems is an emerging idea, the principles behind DOA are not new. Many domains have discovered and are reaping the benefits of applying similar principles. For instance, data-oriented design (DOD) applies many DOA principles on a lower level of abstraction, with claims of significant improvements over analogous \hlr{object-oriented programming} (OOP) solutions. Video game development is one domain where DOD is widespread to improve memory and cache utilisation \cite{acton2014data}. For example, Coherent Labs utilised DOD while creating their game engine Hummingbird \cite{nikolov2018oop} to overcome the performance limitations of solutions based on Chromium and WebKit. DOD helped the developers eliminate cache misses and compiler branch mispredictions, leading to a 6.12x improvement in animation rendering speed. Outside of gaming, Mironov et al.~\cite{Mironov2021comparison} utilised DOD to improve the performance of a trading strategy backtesting utility. They improved the parallelisation opportunities and increased the performance speed to 66\%. These works illustrate that DOA-like principles, especially around data prioritisation, emerge at a different level of abstraction with different motivations and nevertheless bring significant benefits to diverse practitioners. \hlb{Our work extends these previous efforts by reviewing DOA principles in the context of ML-based systems at deployment.}

\hlb{A growing body of literature focuses on surveying the application of AI. Table~\ref{table:related-work-comparison} shows how our work compares against these surveys. One category of current surveys presents use cases of AI algorithms that solve problems in specific domains. They report the challenges these algorithms face at deployment.} For example, Cai et al.~\cite{cai2019survey} survey multimodal ML techniques applied to the healthcare domain. They review systems for disease analysis, triage, diagnosis, and treatment. Bohg et al.~\cite{bohg2013data} survey data-driven methodologies for robot grasping. They focus on techniques based on object recognition and pose estimation for known, familiar, and unknown objects. The review compares data-driven methodologies and analytics approaches. Qin et al.~\cite{qin2012survey} analyse data-driven methods for industrial fault detection and diagnosis. This survey focuses on fault detectability and identifiability methods for industrial processes with different \hlr{complexities and sizes}. Wong et al.~\cite{Wong2020TheRE} present the challenges of applying deep learning (DL) techniques in radio frequency applications. This work reviews DL applications in the radio frequency domain from the perspective of data trust, security, and hardware/software issues for DL deployment in real-world wireless communication applications. Joshi et al.~\cite{Joshi2022} \hlr{survey} the deployment of deep learning approaches at the edge. Their review presents architectures of deep learning models, enabling technologies, and adaptation techniques. Their work also describes different metrics for designing and evaluating deep learning models at the edge. \hlb{Our survey is domain-agnostic and focuses on papers that report ML-based systems at deployment. In addition, we extend specific domain surveys by analysing the design decisions and software architectural patterns behind the systems the authors report.}

\hlb{Another category of papers reviews software engineering patterns for ML applications}. Since DOA is a general architectural style that builds on multiple design patterns, our survey \hl{closely relates} to such existing works. Muccini and Vaidhyanathan \cite{Muccini2021Software} emphasise the importance of research on architectural patterns for ML-based systems and provide high-level research directions. Washizaki et al. \cite{washizaki2020machine} focus on a more in-depth discussion of the patterns for building ML-based systems by reviewing academic literature and engineering blog posts to identify 33 patterns. DOA directly applies or encourages many of these patterns (e.g., ``Kappa Architecture'' and \hlr{``Modularisation} of ML Components''). Chai et al. surveyed challenges that arise at different stages of ML pipelines and acknowledged data management as one of the crucial challenges for ensuring the high quality of a pipeline \cite{chai2022data}. Giray \cite{giray2021software} performed a systematic analysis of software engineering practices for ML-based systems. Heck \cite{heck2024data} conducted a mapping study on data engineering for AI systems, briefly summarising relevant technical solutions and architectural styles. At a high level, these works describe multiple architectural styles, patterns and paradigms for designing and developing ML-based systems. They attempt to deepen the community's understanding of modern software engineering for ML. \hlb{We provide further insights by exclusively reviewing ML-based systems from a data-oriented perspective (i.e., data focus), which is the crucial and distinct aspect of modern and future data-driven systems. Moreover, none of the papers in the related work perform a detailed study of a single architectural style or pattern in the context of ML-based systems. In contrast, our work surveys the adoption level of DOA when deploying ML-based systems. We define the DOA core principles, compare them against ML challenges at deployment, and provide advice on applying DOA in practice.} \hlg{We quantify the adoption of DOA principles among already existing ML-based systems by conducting a deep architectural analysis of these works. Our long-term goal is to unify multiple approaches for deploying ML-based systems under the DOA architectural style and to increase its recognition in the software architecture community.}

\begin{table}
\caption{Related work compared against this survey. \hlr{Our work is the only survey that measures an architectural style adoption (i.e., DOA) in the context of ML-based systems at deployment.}}
\label{table:related-work-comparison}
\begin{tabular}{|l|l|l|l|l|l|}
\hline
\textbf{Related work} & \textbf{Domain} & \thead{Focus on\\Architecture\\Style} & \thead{Focus on\\Deployment} & \thead{Focus on\\Data} & \thead{Focus on\\Adoption}
\\ \hline
Cai et al. \cite{cai2019survey} & Healthcare & No & Yes & No & No
\\ \hline
Bohg et al. \cite{bohg2013data} & Robotics & No & Yes & No & No
\\ \hline
Qin et al. \cite{qin2012survey} & Industrial processes & No & Yes & No & No
\\ \hline
Wong et al. \cite{Wong2020TheRE} & Radio frequency & No & Yes & No & No
\\ \hline
Joshi et al. \cite{Joshi2022} & Edge computing & No & Yes & No & No
\\ \hline
\makecell[l]{Muccini and\\Vaidhyanathan \cite{Muccini2021Software}} & \makecell[l]{ML software\\architecture} & Yes & Yes & No & No
\\ \hline
Washizaki et al. \cite{washizaki2020machine} & ML engineering & Yes & No & No & No
\\ \hline
Chai et al. \cite{chai2022data} & ML pipelines & Yes & No & No & No
\\ \hline
Giray \cite{giray2021software} & ML engineering & Yes & Yes & No & No
\\ \hline
Heck \cite{heck2024data} & Data engineering & No & Yes & Yes & Yes
\\ \hline
This work & ML-based Systems & Yes & Yes & Yes & Yes
\\ \hline
\end{tabular}
\end{table}

\section{Data-Oriented Architectures and ML Challenges}
\label{sec:doa-principles}

\begin{figure}[t!]
	\centering
	\includegraphics[width=\textwidth]{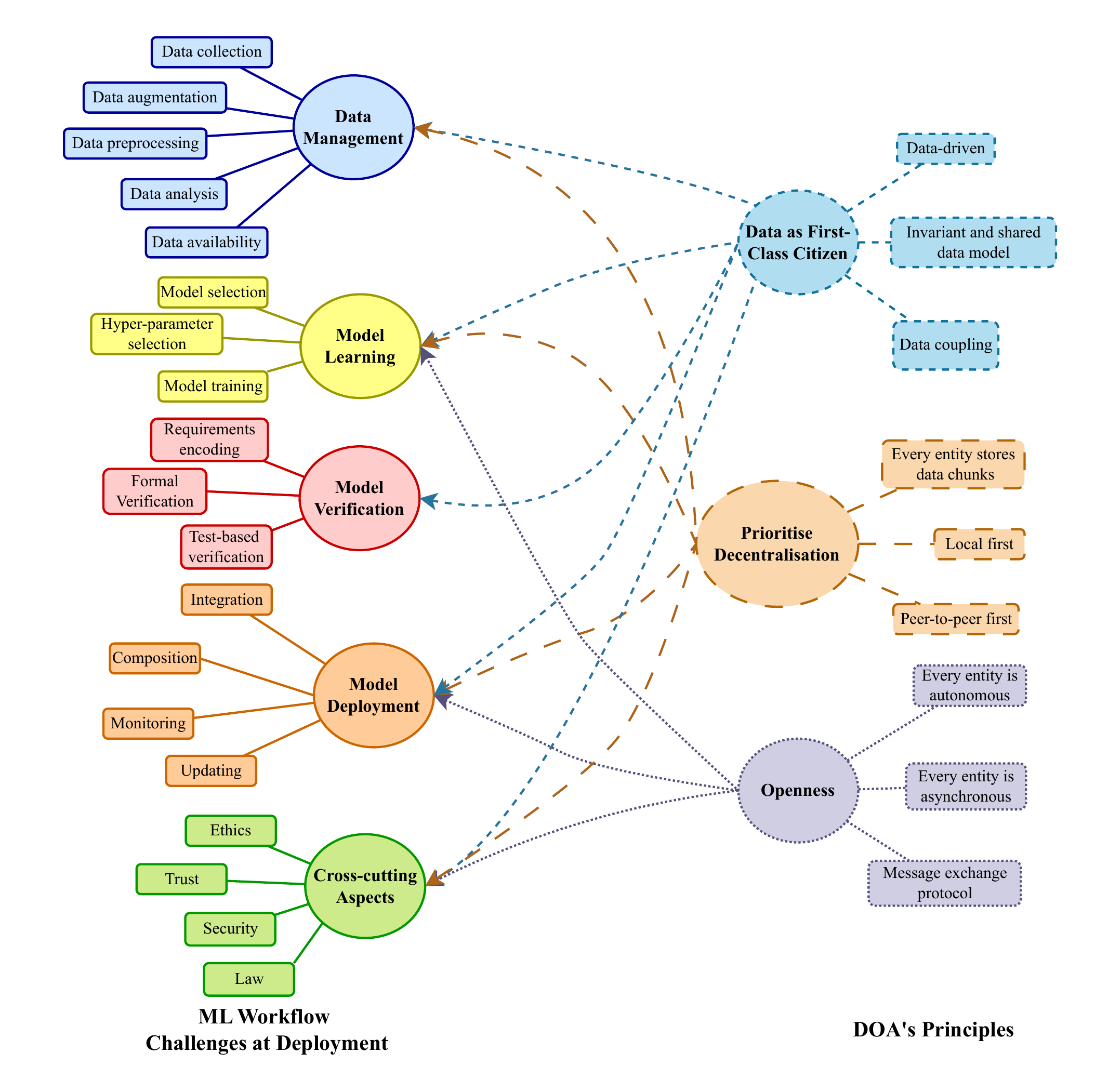}
	\caption{Map between ML Workflow Challenges at Deployment~\cite{paleyes2022challenges} and DOA principles. The left side shows the ML challenges at deployment, and the right side shows the DOA principles. The links between them represent which principles support addressing the respective challenges.}
	\label{fig:map-challenges-principles}
    \vspace{-5mm}
\end{figure}

\hl{Service-oriented architecture (SOA) is the prevalent architectural style to design and implement modern software.} It defines services as individual components that expose their functionalities through interfaces (APIs) and interact via calls to these interfaces~\cite{stopford2016data}. \hlr{Modularity enables the assignment of systems components to different teams, allows flexible scaling, and facilitates maintainability.} Independent departments in the organisation focus on specific services instead of the whole system because of the well-defined interfaces between the parts of the system~\cite{microservices2016dividends,microservices2019challenges}. SOAs usually deploy services and microservices in cloud platforms for high availability and robustness. Flexible and automatic management of resources (e.g., replicas, load-balancers, etc.) enables systems to be fault-tolerant and guarantee the quality of service despite the amount of data to be processed~\cite{microservices2021deployment}. Maintainability, high availability, and scalability are essential requirements that ML-based systems must satisfy when deployed in the real world. Service abstractions and cloud deployments enable current system architectures to meet these requirements. However, real-world environments pose additional requirements to ML-based systems, challenging SOA. Data monitoring and transparency are challenging because services hide data behind their interfaces (``The Data Dichotomy'', \cite{stopford2016data}). Service interactions based on interfaces prioritise control flow and make data exchanged during the communication ephemeral~\cite{paleyes2022fbpsoa}. Cloud platforms are expensive and take ownership of end users' data. The physical distance between end users and cloud servers impacts the end-to-end latency of critical applications supported by AI (e.g., real-time health diagnosis)~\cite{cabrera2023maaco}. Cloud resources are usually expensive, failing to satisfy resource constraint requirements when there are budget constraints in the real-world systems~\cite{schwartz2020greenai,cloud2021limitations}. Cloud deployments imply the change of data ownership from end users to service providers. End users are unaware of where and how providers process their sensitive data and the associated security risks. Data ownership, security, accountability, and trust requirements are hard to satisfy in such deployments~\cite{Kleppmann2019localfirst,akbar2023security,cobbe2023accountability,dantonoli2024large}.

Data-Oriented Architecture (DOA) is an emerging \hl{architectural style} that can help us overcome these challenges when \hl{implementing and deploying} ML-based systems. DOA proposes a set of principles for implementing data-oriented software: considering data as a first-class citizen, decentralisation as a priority, and openness~\cite{joshi2007data,schuler2015data,vorhemus2017data, Kleppmann2019localfirst}. Figure~\ref{fig:map-challenges-principles} maps ML deployment challenges and DOA principles addressing them. The left side of the figure shows the challenges of the ML workflow at deployment discussed by Paleyes et al.~\cite{paleyes2022challenges}, while the right side shows the DOA principles we have extracted from the literature~\cite{joshi2007data,schuler2015data,vorhemus2017data, Kleppmann2019localfirst}. This section introduces each DOA principle and discusses how they support the deployment of ML-based systems in the real world.

\subsection{Data as a First Class Citizen}
\label{sec:datafirst}

Data collection, preprocessing, analysis, and monitoring tasks are challenging for systems deployed in the real world. Such environments usually generate high volumes of variable data, which needs to be processed in real time to create value. These requirements come from the properties of the data real-world systems must handle, which are also known as "the five Vs" of big data: volume, variability, velocity, veracity, and value~\cite{nazabal2020data}. Such data properties and requirements become more critical for ML-based systems implementation because they are inherently data-driven. The outputs and overall performance of ML-based systems depend on the quality of the data that flows through their components~\cite{fisler2021datacentric}. Practitioners must design data-driven systems focusing on the specific data to be handled and processed. These systems must monitor their data to evaluate performance at runtime, identify or predict failures, and trigger maintenance or adaptation operations. Architectural styles oriented to individual services encapsulate atomic capabilities and interact through their interfaces, falling into "The Data Dichotomy". These architectures hide their data behind service interfaces, but data management requires exposing data~\cite{stopford2016data}. The data dichotomy does not suit data management tasks because additional efforts are needed to access the data that flows through the system. Such data unavailability complicates systems' monitoring and transparency and the related tasks~\cite{zittrain2022debt,cobbe2023accountability}. In the ML context, it means that systems' data unavailability challenges the selection, parametrisation, training, testing, verification, integration, monitoring, composition, and updating of the learning model components as well as systems' data transparency requirements~\cite{paleyes2022challenges}.

DOA proposes to treat the \textbf{data as a first-class citizen} of a software system, understanding data as the common denominator between disparate components~\cite{joshi2007data}. \hlr{The data in a DOA-based system is primary, and the operations on the data are secondary.} This principle makes systems \textit{data-driven} by design, which matches the nature of ML-based components. DOA-based systems rely on a \textit{shared data model}, which is processed and nourished by multiple system components~\cite{schuler2015data}. The shared data model is a single data structure equivalent to the one that data engineers build into the data management stage of ML workflows. The key difference is that the system automatically creates this data model from its components' interactions. Systems components do not expose any APIs and interact via data mediums, where input is listened to, and output is offloaded~\cite{schuler2015data}. Such interactions enable DOA-based systems to achieve \textit{data coupling}, considered the loosest form of coupling~\cite{OLSSON753212,offutt1993software}. The shared data model stores the history of the system's state during its whole life cycle. Systems' data is fully available, describing the systems' current and past states. Data availability facilitates data management, learning model management, systems monitoring, failure detection, and adaptation tasks. Components' behaviour is programmatically observable, traceable, and auditable. Such transparency helps the responsible \hl{design and implementation} of ML-based systems to resolve questions regarding accountability, trust, and traceability.

\subsection{Prioritise Decentralisation}
\label{sec:decentralise}

Big tech companies drove recent ML breakthroughs thanks to the increasingly powerful hardware and the efforts in building massive data sets~\cite{lecun2015deep,dovsilovic2018explainable}. These enablers are not always present in systems deployed in the real world. Smaller organisations are using ML to solve challenging problems (e.g. sustainable farming, climate change, etc.) with limited budgets and resources~\footnote{An example of such an organisation is Data Science Africa (DSA): \url{http://www.datascienceafrica.org/aboutus/}}. Most organisations worldwide cannot afford expensive cloud computing resources or the effort of building and maintaining massive learning models. Such a reality threatens the practical and democratic adoption of ML models, as different stages of their workflow are computationally expensive. For example, the training stage is an iterative process that solves an optimisation problem to find learning model parameters. Complex models (e.g., large language models)  make this process computationally expensive when, for example, they learn from non-quadratic, non-convex, and high-dimensional data sets~\cite{judd1990neural,orr2003neural,goodfellow2017deep}. Hyperparameters improve the efficiency of the training process and the accuracy of the learning models~\cite{goodfellow2014qualitatively}. However, selecting these hyperparameters is also a resource-demanding optimisation problem~\cite{paine2020hyperparameter,bischl2021hyperparameter}. In addition, expensive and physically centralised deployments can fail to meet low-latency requirements from critical applications. The physical distance between end users and data centres impacts systems' end-to-end response time~\cite{cabrera2023maaco}. Data ownership, security, and trust requirements are also affected in centralised systems. The ownership of the data changes from end users to service providers, which can entail privacy and security issues as end users are unaware of where and how their data is stored, processed, and used~\cite{akbar2023security}. Recent research shows that Large Language Models (LLMs) create risks of leaking sensitive information from the model's training data. Adversarial attacks or prompt engineering strategies can extract personal information from LLMs~\cite{dantonoli2024large}.

The simplest way to deliver a shared data model (Principle~\ref{sec:datafirst}) would seem to be to centralise it, but, in practice, scalability, resource constraints, low latency, and security requirements mean that in DOA we \textbf{prioritise decentralisation}~\cite{vorhemus2017data, Kleppmann2019localfirst}. Such decentralisation should be logical and physical. Logical decentralisation enables organisations to scale when developing ML-based systems, as different development teams focus on smaller systems' components. Logical decentralisation is a way to achieve a clear separation of concerns, avoid central bottlenecks in the computational process, and increase flexibility in system development. Physical decentralisation enables the deployment of ML models in constrained environments without expensive computational resources and improves privacy and ownership of data~\cite{Kleppmann2019localfirst}. Practitioners should deploy components of ML-based systems as decentralised \textit{entities that store data chunks} of the shared information model described in the previous principle. These entities first perform their operations with their local resources (i.e. \textit{local first},~\cite{Kleppmann2019localfirst}). \hlr{If local resources (i.e., data, computing time, or storage) are insufficient, entities can connect temporarily with other participants to share resources.} Entities first scan their local environment for potential resources they need. They prioritise interactions with nodes in the close vicinity to share or ask for data and computing resources (i.e., \textit{peer-to-peer first}). Centralised servers are fallback mechanisms~\cite{vorhemus2017data}. This principle facilitates data management, enabling the system's data replication by design because different entities can store the same data chunk. Such replication provides data availability because if one entity fails, its information is not lost. Similarly, replication provides scalability as different entities can respond to concurrent data requests~\cite{vorhemus2017data}. Prioritised decentralisation also alleviates the high demand for resources of ML-based systems and reduces the inference latency. ML-based systems can perform their data-related tasks (e.g., ML model training) in smaller data sets partitioned by design, and trained models deployed closer to end users to provide faster responses~\cite{cabrera2021rl}. In addition, decentralisation creates a flexible ecosystem where we can use resources from different devices on demand. \hlr{This DOA principle advocates for a more sustainable and democratic approach that prioritises the computational power available in everyday devices over expensive cloud resources.} It is important to note that logical or physical decentralisation is not always necessary. For example, there is no need to perform local first computations when centralised computational resources are available. However, data replication, partitioned data sets, and flexible resource management are DOA-enabled properties that can still benefit even partial decentralisation of ML-based systems.

\subsection{Openness}
\label{sec:openness}

Developing ML components often involves the automation of various stages of data processing, both at the training and inference stages. The amounts of data ML-based systems manage, the complex processes they perform, and the fact that their users are usually experts in domains different from ML (e.g. healthcare, physics, etc.)~\cite{waring2020automated} demand such automation. AutoML is a recent research area that aims to automate the ML training life cycle. AutoML approaches include data augmentation, model selection, and hyperparameter optimisation~\cite{escalante2020automated,vaccaro2021empirical}. Nevertheless, implementing and deploying ML-based systems in real-world environments sets automation requirements for adopting ML models that AutoML does not cover. Real-world environments are dynamic, and deployed systems must respond autonomously to changing goals, variable data, unexpected failures, security threats, complex uncertainties, etc. Autonomous responses are required because practitioners do not have complete control over such deployments~\cite{pahl2023research}, their complexity usually exceeds the limit of human comprehension~\cite{paleyes2023causal}, and human intervention is not feasible~\cite{cabrera2022self}. Current ML-based systems rely on the interaction of heterogeneous components. Static systems architectures usually pre-define the components and devices that compose a system, and third-party entities (e.g., central controllers) usually integrate, orchestrate, and maintain these components to satisfy end-to-end system quality requirements~\cite{cabrera2017middleware}. The scale and dynamic nature of real-world deployments demand autonomous solutions and challenge these static and centralised architectures.

Autonomous and independent software components are the building blocks of systems that can adapt and respond to uncertain environments without human intervention~\cite{weyns2019software}. DOA proposes \textbf{openness} as a principle that enables designing and implementing such autonomous systems' components in open environments where they interact autonomously~\cite{joshi2007data}. DOA represents components as \textit{autonomous} and \textit{asynchronous} entities that communicate with each other using a common \textit{message exchange protocol}. Systems can leverage such environments to integrate ML models because their decentralised components can autonomously perform the integration, composition, monitoring, and adaptation tasks. Asynchronous entities produce their outputs and can subscribe to inputs at any time~\cite{schuler2015data}. These steps are transparent, favouring data trust and traceability. Similarly, the system's components are autonomous in deciding which data to store, which to make public, and which to hide for security and privacy~\cite{vorhemus2017data}. The message exchange protocol between asynchronous components replaces interface dependencies in service-oriented or microservices architectures with asynchronous messages between data producers and consumers. Asynchronous communication protocols enable data coupling, considered the loosest form of coupling, enabling scalability and low latency requirements~\cite{OLSSON753212,offutt1993software}.
\section{Survey Methodology}
\label{sec:survey-method}

\begin{figure}[t]
	\centering
	\includegraphics[width=\textwidth]{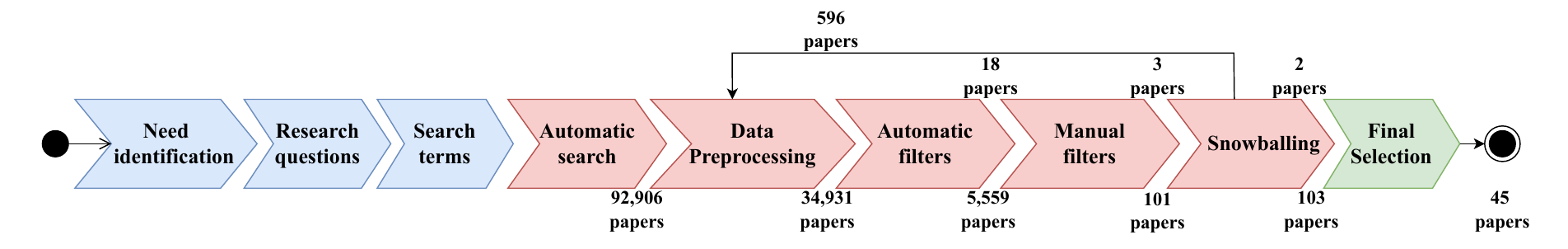}
	\caption{The survey process depicts the steps from the review motivation to the full-text reading of the selected papers. This process relies on the methodology proposed by Kitchenham et al.~\cite{kitchenham2007guidelines, Kitchenham20132049}}
	\label{fig:survey-process}
    \vspace{-5mm}
\end{figure}

\hlb{This survey assesses why, how, and to what extent practitioners have adopted the DOA principles when implementing and deploying ML systems and the motivation and methods behind such adoption.} We survey research works that have used ML to solve problems in different domains and have been deployed and tested in real-world settings. We are particularly interested in works that report the software architectures behind these systems and, if possible, the authors' design decisions toward such architectures.

The selection of relevant works for our survey is not straightforward because many papers that apply ML in different domains have been published in recent years, making manual selection unfeasible. For this reason, we developed a semi-automatic framework based on a well-known methodology for systematic literature reviews (SLRs) in software engineering~\cite{kitchenham2007guidelines, Kitchenham20132049}. The framework is available publicly on GitHub\footnote{Semi-automatic Literature Survey: \url{https://github.com/cabrerac/semi-automatic-literature-survey/tree/doa-survey}} to allow the reproducibility of this work, as well as the reusability of this framework in other surveys. The framework queries the search APIs from different digital libraries to automatically retrieve papers' metadata (e.g., title, abstract, and citations). It then applies syntactic and semantic filters over the retrieved records to reduce the search space. Then, we manually filtered the list of works to select the papers to survey. Figure~\ref{fig:survey-process} depicts the stages of the survey process. It has two principal stages described in the next section.


\begin{table}
\caption{Search query format used to retrieve papers. The query is composed of three search terms in conjunction. The values in the second column replace each search term.}
\label{tab:search-query}
\begin{tabular}{|l|l|}
\hline
\thead{Search Query} & \textless{}search\_term\_1\textgreater{} AND \textless{}search\_term\_2\textgreater{} AND \textless{}search\_term\_3\textgreater{}
\\ \hline
\thead{\textless{}search\_term\_1\textgreater{}} & \makecell[l]{"autonomous vehicle" OR "health" OR "industry" OR "smart cities" OR  "multimedia" OR \\ "science" OR "robotics" OR "oceanology" OR "finance" OR "space" OR "e-commerce"}
\\ \hline
\thead{\textless{}search\_term\_2\textgreater{}} & "machine learning"
\\ \hline
\thead{\textless{}search\_term\_3\textgreater{}} & "real world" AND "deploy"
\\ \hline
\end{tabular}
\vspace{-3mm}
\end{table}

\begin{table}
\caption{Synonyms to extend queries. Search terms in the Word column are expanded using their respective synonyms.}
\label{tab:synonyms}
\begin{tabular}{|l|l|}
\hline
\thead{Word} & \thead{Synonyms}
\\ \hline
"health" & "healthcare", "health care", "healthcare", "medicine", "medical", "diagnosis"
\\ \hline
"industry" & "industry 4", "manufacture", "manufacturing", "factory", "manufactory", "industrial"
\\ \hline
"smart cities" & \makecell[l]{"sustainable city", "smart city", "digital city", "urban", "city", "cities", "mobility", \\ "transport", "transportation system"}
\\ \hline
"multimedia" & \makecell[l]{"virtual reality", "augmented reality", "3D", "digital twin", "video games", "video", \\ "image recognition", "audio", "speech recognition", "speech"}
\\ \hline
"science" & \makecell[l]{"physics", "psychology ", "chemistry", "biology", "geology", "social", "maths",\\"materials", "astronomy", "climatology", "oceanology", "space"}
\\ \hline
"autonomous vehicle" & \makecell[l]{"self-driving vehicle", "self-driving car", "autonomous car", "driverless car", \\"driverless vehicle", "unmanned car", "unmanned vehicle", "unmanned aerial vehicle"}
\\ \hline
"networking" & "computer network", "intranet", "internet", "world wide web"
\\ \hline
"e-commerce" & "marketplace", "electronic commerce", "shopping", "buying"
\\ \hline
"robotics"  & "robot"
\\ \hline
"finance" & "banking"
\\ \hline
"machine learning" & \makecell[l]{"ML", "deep learning", "neural network", "reinforcement learning", \\"supervised learning", "unsupervised learning", "artificial intelligence", "AI"}
\\ \hline
"deploy" & "deployment", "deployed", "implemented", "implementation", "software"
\\ \hline
"real world" & "reality", "real", "physical world"
\\ \hline
\end{tabular}
\vspace{-3mm}
\end{table}

\subsection{\textbf{Planning Stage}}

The first stage consists of the review plan definition as follows:

\begin{itemize}
    
    \item [a.]\textit{Need identification:} Section~\ref{sec:doa-principles} introduced the DOA principles and how these can support practitioners when deploying ML in the real world. \hlr{Despite these potential benefits and several surveys on ML and its applications (see Section~\ref{sec:related-work}), it remains unclear to what extent, why, and how practitioners have adopted these principles in deploying ML-based systems. Practitioners usually follow DOA principles implicitly, creating a knowledge gap around the DOA style and limiting the implementation of ML-based systems. We close such a gap by surveying deployed ML-based systems from a DOA perspective to identify the design decisions, implementation practices, and deployment strategies practitioners followed.}
    
    \item [b.]\textit{Research questions:} \hlb{The main research question we want to answer with this survey is: \textit{To what extent and how have researchers adopted the principles of Data-Oriented Architecture (DOA) in the implementation and deployment of modern ML-based systems, and what are the motivations behind such adoption?} We split this question into three complementary research questions to determine the level and way of adoption of each DOA principle in deployed ML-based systems: \textbf{data as a first-class citizen (RQ1)}, \textbf{prioritise decentralisation (RQ2)}, and \textbf{openness (RQ3)}.} The answers to these questions will enable us to identify the research gaps and inform the development of the next generation of DOAs. Our long-term goal is to establish DOA as a mature and competitive \hl{architectural style} for designing, developing, implementing, deploying, monitoring, and adapting ML-based systems.

    \item [c.]\textit{Search terms:} \hlb{We want to search for papers presenting ML-based systems deployed in real-world environments to answer the research questions above.} Table~\ref{tab:search-query} shows the query format and the search terms we use to retrieve such papers. The query has three search terms in conjunction (i.e., AND operator). The first term refers to popular domains that have applied ML. The second term filters papers that apply machine learning in these domains, and the third term filters papers that deploy their solution in the real world. Search engines in scientific databases use different matching algorithms. Some engines search for exact words in the papers' attributes (e.g., title or abstract), which can be too restrictive. We expand these queries by including synonyms for the search terms (Table~\ref{tab:synonyms}). Synonyms extend queries using inclusive disjunction (i.e. OR operator).  
    
    \item [d.]\textit{Source selection:} We search for papers in the most popular scientific repositories using their APIs. They are IEEEXplore\footnote{IEEEXplore API: \url{https://developer.ieee.org/}}, Springer Nature\footnote{Springer Nature API: \url{https://dev.springernature.com/}}, Scopus\footnote{Scopus API: \url{https://www.elsevier.com/solutions/sciencedirect/librarian-resource-center/api}}, Semantic Scholar\footnote{Semantic Scholar: \url{https://www.semanticscholar.org/product/api}}, CORE\footnote{CORE API: \url{https://core.ac.uk/services/api}}, and ArXiv\footnote{ArXiv API: \url{https://arxiv.org/help/api/}}. Some repositories, such as the ACM digital library, could not be used as they do not provide an API to query. Nevertheless, we are confident in the sufficient coverage of our search because of the significant overlap with other sources (papers can be published in multiple libraries or indexed by meta-repositories such as Semantic Scholar).
\end{itemize}

\begin{table}
\caption{We used the following categories and keywords for the Lbl2Vec algorithm~\cite{lbl2vec2021}.}
\label{tab:categories}
\begin{tabular}{|l|l|}
\hline
\multicolumn{1}{|c|}{\textbf{Category}} & \multicolumn{1}{c|}{\textbf{Keywords}}                                                                                                                           \\ \hline
"system"                                & "architecture", "framework", "platform", "tool",  "prototype".                                                                                                   \\ \hline
"software"                              & \begin{tabular}[c]{@{}l@{}}"develop", "engineering", "methodology", "architecture", "design", \\ "implementation", "open", "source", "application".\end{tabular} \\ \hline
"deploy"                                & \begin{tabular}[c]{@{}l@{}}"production", "real", "world", "embedded", "physical", "cloud", "edge",  \\ "infrastructure".\end{tabular}                            \\ \hline
"simulation"                            & "synthetic", "simulate".                                                                                                                                         \\ \hline
\end{tabular}
\end{table}

\subsection{\textbf{Conducting Stage}}

The second stage consists of the review execution as follows:

\begin{table}
\caption{Number of papers selected at each stage of the system literature review process from each scientific repository.}
\label{table:selected_papers}
\begin{tabular}{|l|l|l|l|l|l|l|l|}
\hline
\textbf{Repository} & \thead{Automatic\\Search} & \thead{Pre-\\processing} & \thead{Automatic\\Filters} & \thead{Manual\\Filters} & \thead{Snow-\\balling} & \thead{Final\\Selection}
\\ \hline
IEEEXplore & 12,467 & 9,084 & 2,557 & 58 & 58 & 21
\\ \hline
Springer Nature & 58,199 & 15,617 & 463 & 7 & 7 & 2
\\ \hline
Scopus & 4,537 & 2,239 & 727 & 10 & 10 & 8
\\ \hline
Semantic Scholar & 12,308 & 5,225 & 638 & 6 & 8 & 5
\\ \hline
CORE & 1,917 & 386 & 251 & 2 & 2 & 2
\\ \hline
ArXiv & 3,480 & 2,374 & 923 & 18 & 18 & 7
\\ \hline
\textbf{Total Papers} & \textbf{92,908} & \textbf{34,931} & \textbf{5,559} & \textbf{101} & \textbf{103} & \textbf{45}
\\ \hline
\end{tabular}
\end{table}

\begin{itemize}
    \item [a.]\textit{Automatic search:} We implemented clients that consume the APIs exposed by the selected repositories as part of our semi-automatic framework. In this step, each client creates the query in the format the respective API expects, submits the request, and stores the search results in separate CSV files. The search results are the metadata of the retrieved papers (e.g., title, abstract, publication data).
    \item [b.]\textit{Preprocessing of Retrieved Data:} Each API provides papers' metadata in a different format. There are duplicated papers between repositories, and some records can be incomplete (e.g., a paper missing an abstract). The preprocessing step prepares the data for the following steps in our semi-automated framework. It joins the papers' metadata in a single file, cleaning the data and removing repeated and incomplete records. The process selected a total of 34,931 works after this step.
    \item [c.]\textit{Automatic Filtering:} \hlb{All the data of the retrieved papers are stored in a single file after the preprocessing step. However, the number of papers is still too large for manual processing. We reduce the search space by applying two filters. A syntactic filter selects the papers that talk in the abstract about real-world deployments \hl{according to our definition of "ML systems in the real world"}. In particular, this filter searches for the "real world" and "deploy" words and their synonyms (Table~\ref{tab:synonyms}) in the papers' abstracts. We classified the selected papers into four categories after the syntactic filtering. The first category includes the papers that present architectures of deployed ML-based systems, the second category groups papers that present software engineering approaches to build ML-based systems in practice, and the third category represents papers that show physical implementations (e.g., edge architectures) of ML-based systems with a special focus on the infrastructure. The final category includes papers that evaluate ML algorithms and systems based on synthetic data and simulated environments. We used an unsupervised-learning algorithm, Lbl2Vec, to semantically classify the selected papers in these four groups following the work proposed by Schopf et al.~\cite{lbl2vec2021}. Lbl2Vec requires as inputs a set of texts to classify (i.e., selected papers' abstracts) and a set of predefined categories (Table~\ref{tab:categories}). The algorithm assigns papers to the most relevant category. We use this semantic classification to select the papers that belong to the first three categories (i.e., system, software, and deploy). These filters produced a total of 5,559 papers.}
    \item [d.]\textit{Manual filtering:} \hlb{The syntactic filters in the previous step reduce the set of papers to a more feasible number to be manually explored. Our framework in this step shows the paper information to the user in a centralised interface where papers are selected as included or excluded. Such manual selection has two stages following the methodology defined by Kitchenham et al.~\cite{kitchenham2007guidelines, Kitchenham20132049}. We select the papers by reading their abstracts in the first stage. Then, we filter the works by skimming the full text. The skimming process includes assessing the papers' quality. According to the goal of our survey, we focused on papers whose content introduces the software architectures behind ML-based systems and reports the results of their deployment in the real world. We discarded the works that did not present these two attributes. One researcher performed the manual filters that produced 101 papers.}
    \item [e.]\textit{Snowballing:} We use the API from Semantic Scholar to retrieve metadata of the papers that cite the selected works from the previous stage. Our tool preprocesses and filters (i.e., syntactically and semantically) the citing papers and adds the ones that pass these filters to the final set of selected papers. This process resulted in 2 more papers added to the selection, resulting in 103 papers for full-text reading.
    \item [f.]\textit{Final Selection:} Two researchers read the 103 papers, selected 45 papers, and extracted data from these to answer our survey research questions. We agreed on the selection criteria based on two aspects. First, we considered the papers' relevance by identifying if they introduce the architectures behind the ML-based systems and report results from real-world deployments. Second, we use the DOA principles definition to determine to what extent these papers adopt the DOA principles (Section~\ref{sec:doa-principles}). We used a data collection spreadsheet\footnote{Data collection spreadsheet: \url{https://github.com/cabrerac/semi-automatic-literature-survey/blob/doa-survey/papers/doas_search/2022_08_04/doas-survey.xlsx}} to analyse and extract the data from the 45 selected papers. The spreadsheet has each paper's metadata (i.e., ID, URL, and title) and columns each researcher must fill out according to the paper analysis. We use the "Quick Summary" column to describe the research work, its domain, problem, solution, and reasons for inclusion or exclusion. The "Real-World Deployment" column describes how authors deploy the selected works in the real world. We use the "Data as First Class Citizen" (i.e., RQ1), "Decentralised Architecture (i.e., RQ2), and "Open Architecture"  (i.e., RQ3) columns to describe the extent to which and how these papers follow the DOA principles. We use the "Data Flow Design" and "Data Coupling" to evaluate if the authors' design systems with a focus on data flowing between their components (i.e., data-first) and if such components communicate through data mediums (i.e., decentralisation and openness). For each of these columns, we defined three possible outcomes:

    \begin{itemize}
        \item \textbf{"YES"} meaning the reviewed paper fully follows the evaluated principle. For example, a system fully follows the data as a first-class citizen principle if the system is data-driven, its components share a data model, and they communicate through such a data model (i.e., data coupling).
        \item \textbf{"PARTIAL"} meaning the reviewed paper partially follows the evaluated principle. For example, a system partially prioritises decentralisation if it stores and processes data chunks locally but does not have peer-to-peer communication with components in the same layer.
        \item \textbf{"NO"} meaning the reviewed paper does not follow the evaluated principle in any way. For example, a system does not follow the openness principle when it does not implement autonomous and asynchronous entities exchanging messages between them.
    \end{itemize}
    
    Each researcher justifies the outcome for each column and paper in the data collection spreadsheet. In addition, the spreadsheet also has a column "Comments" to annotate extra information researchers find relevant for our study. We addressed selection and data extraction conflicts in group discussions. Section~\ref{sec:doa-survey} presents the survey of the 45 papers using a descriptive and quantitative synthesis of the results we extracted in the spreadsheet.
    
    \item [g.]\textit{Exclusion criteria:} We exclude the following works during the different stages of our semi-automatic process:
        \begin{itemize}
            \item \hl{Papers that introduce ML-based systems without deployment and evaluation in the real world or a production environment.}
            \item \hl{Papers that present ML-based systems but do not report their software architecture.}
            \item \hl{Papers that report experiments on synthetic data or simulated environments where ML-based systems do not interact with actual data, physical entities, or people.}
            \item \hl{Papers presenting isolated ML algorithms not part of larger systems.}
            \item Papers with missing metadata that our framework cannot analyse (e.g., a paper without an abstract).
            \item Duplicated papers.
            \item Papers not written in English.
            \item Survey and review papers.
            \item Thesis and report documents.
        \end{itemize}
        \hlr{We use the exclusion criteria to distinguish ML-based systems deployed in the real world from ML systems or algorithms developed and tested in controlled and isolated environments.}
\end{itemize}
\section{ML-based Systems Survey from DOA Principles Perspective}
\label{sec:doa-survey}

\begin{table}
\caption{All reviewed papers in our survey. This table shows the extent to which and in what papers practitioners adopt (fully or partially) each of the DOA sub-principles discussed in Section \ref{sec:doa-principles}.}
\resizebox{0.7\textwidth}{!}{%
\label{tab:survey}
\begin{tabular}{l|lll|lll|lll|}
\cline{2-10} & 
    \multicolumn{3}{c|}{\textbf{\begin{tabular}[c]{@{}c@{}}Data as a First \\ Class Citizen\end{tabular}}} &
    \multicolumn{3}{c|}{\textbf{\begin{tabular}[c]{@{}c@{}}Prioritise \\ Decentralisation\end{tabular}}} &
    \multicolumn{3}{c|}{\textbf{\begin{tabular}[c]{@{}c@{}}Openness\end{tabular}}}
\\ \hline
    \multicolumn{1}{|c|}{\textbf{\begin{tabular}[c]{@{}c@{}}Research\\ work\end{tabular}}}  & 
    \multicolumn{1}{c|}{\rotatebox[origin=c]{90}{\textbf{\begin{tabular}[c]{@{}c@{}}Data\\ driven\end{tabular}}}} & 
    \multicolumn{1}{c|}{\rotatebox[origin=c]{90}{\textbf{\begin{tabular}[c]{@{}c@{}}Shared \\ data mode\end{tabular}}}} & 
    \multicolumn{1}{c|}{\rotatebox[origin=c]{90}{\textbf{\begin{tabular}[c]{@{}c@{}}Data \\ coupling\end{tabular}}}} & 
    \multicolumn{1}{c|}{\rotatebox[origin=c]{90}{\textbf{\begin{tabular}[c]{@{}c@{}}Local\\ data chunks\end{tabular}}}} & 
    \multicolumn{1}{c|}{\rotatebox[origin=c]{90}{\textbf{\begin{tabular}[c]{@{}c@{}}Local\\ first\end{tabular}}}} & 
    \multicolumn{1}{c|}{\rotatebox[origin=c]{90}{\textbf{\begin{tabular}[c]{@{}c@{}}Peer-to-peer\\ first\end{tabular}}}} & 
    \multicolumn{1}{c|}{\rotatebox[origin=c]{90}{\textbf{\begin{tabular}[c]{@{}c@{}}Autonomous\\ entities\end{tabular}}}} & 
    \multicolumn{1}{c|}{\rotatebox[origin=c]{90}{\textbf{\begin{tabular}[c]{@{}c@{}}Asynchronous\\ entities\end{tabular}}}} & 
    \multicolumn{1}{c|}{\rotatebox[origin=c]{90}{\textbf{\begin{tabular}[c]{@{}c@{}}Message exchange\\ protocol\end{tabular}}}}
\\ \hline
    \multicolumn{1}{|l|}{Lebofsky et al.~\cite{lebofsky2019breakthrough}} & 
    
    \multicolumn{1}{c|}{\checkmark} & 
    \multicolumn{1}{c|}{\checkmark} & 
    \multicolumn{1}{c|}{\checkmark} & 
    
    \multicolumn{1}{c|}{\checkmark} &
    \multicolumn{1}{c|}{\checkmark} & 
    \multicolumn{1}{c|}{\checkmark} & 
    
    \multicolumn{1}{c|}{\checkmark} & 
    \multicolumn{1}{c|}{\checkmark} & 
    \multicolumn{1}{c|}{\checkmark}
\\ \hline
    \multicolumn{1}{|l|}{Herrero et al.~\cite{herrero2022i40}} & 
    
    \multicolumn{1}{c|}{\checkmark} & 
    \multicolumn{1}{c|}{\checkmark} & 
    \multicolumn{1}{c|}{\checkmark} & 
    
    \multicolumn{1}{c|}{\checkmark} &
    \multicolumn{1}{c|}{\checkmark} & 
    \multicolumn{1}{c|}{\checkmark} & 
    
    \multicolumn{1}{c|}{\checkmark} & 
    \multicolumn{1}{c|}{\checkmark} & 
    \multicolumn{1}{c|}{\checkmark}
\\ \hline
    \multicolumn{1}{|l|}{Zhang et al.~\cite{zhang2016emotion}} & 
    
    \multicolumn{1}{c|}{\checkmark} & 
    \multicolumn{1}{c|}{\checkmark} & 
    \multicolumn{1}{c|}{\checkmark} & 
    
    \multicolumn{1}{c|}{\checkmark} & 
    \multicolumn{1}{c|}{\checkmark} & 
    \multicolumn{1}{c|}{\checkmark} & 
    
    \multicolumn{1}{c|}{\checkmark} & 
    \multicolumn{1}{c|}{\checkmark} & 
    \multicolumn{1}{c|}{\checkmark}
\\ \hline
    \multicolumn{1}{|l|}{Dai et al.~\cite{dai2019bigdl}} & 
    
    \multicolumn{1}{c|}{\checkmark} & 
    \multicolumn{1}{c|}{\checkmark} & 
    \multicolumn{1}{c|}{\checkmark} & 
    
    \multicolumn{1}{c|}{\checkmark} &
    \multicolumn{1}{c|}{\checkmark} & 
    \multicolumn{1}{c|}{\checkmark} & 
    
    \multicolumn{1}{c|}{\checkmark} & 
    \multicolumn{1}{c|}{\checkmark} & 
    \multicolumn{1}{c|}{\checkmark}
\\ \hline
    \multicolumn{1}{|l|}{Junchen et al.~\cite{Junchen2017}} & 
    
    \multicolumn{1}{c|}{\checkmark} & 
    \multicolumn{1}{c|}{\checkmark} & 
    \multicolumn{1}{c|}{\checkmark} & 
    
    \multicolumn{1}{c|}{--} & 
    \multicolumn{1}{c|}{\checkmark} & 
    \multicolumn{1}{c|}{\checkmark} & 
    
    \multicolumn{1}{c|}{\checkmark} & 
    \multicolumn{1}{c|}{\checkmark} & 
    \multicolumn{1}{c|}{\checkmark}   
\\ \hline
    \multicolumn{1}{|l|}{Karageorgou et al.~\cite{karageorgou2020sentiment}} & 
    
    \multicolumn{1}{c|}{\checkmark} & 
    \multicolumn{1}{c|}{--} & 
    \multicolumn{1}{c|}{\checkmark} & 
    
    \multicolumn{1}{c|}{\checkmark} &
    \multicolumn{1}{c|}{\checkmark} & 
    \multicolumn{1}{c|}{\checkmark} & 
    
    \multicolumn{1}{c|}{\checkmark} & 
    \multicolumn{1}{c|}{\checkmark} & 
    \multicolumn{1}{c|}{\checkmark}

\\ \hline
    \multicolumn{1}{|l|}{Shan et al.~\cite{shan2022poligraph}} & 
    
    \multicolumn{1}{c|}{\checkmark} & 
    \multicolumn{1}{c|}{\checkmark} & 
    \multicolumn{1}{c|}{\checkmark} & 
    
    \multicolumn{1}{c|}{\checkmark} & 
    \multicolumn{1}{c|}{\checkmark} & 
    \multicolumn{1}{c|}{\checkmark} & 
    
    \multicolumn{1}{c|}{\checkmark} & 
    \multicolumn{1}{c|}{--} & 
    \multicolumn{1}{c|}{--}
\\ \hline
    \multicolumn{1}{|l|}{Zhang et al.~\cite{zhang202148learningadd}} & 
    
    \multicolumn{1}{c|}{\checkmark} & 
    \multicolumn{1}{c|}{\checkmark} & 
    \multicolumn{1}{c|}{\cellcolor{gray!25}} & 
    
    \multicolumn{1}{c|}{\checkmark} & 
    \multicolumn{1}{c|}{\checkmark} & 
    \multicolumn{1}{c|}{\checkmark} & 
    
    \multicolumn{1}{c|}{--} & 
    \multicolumn{1}{c|}{--} & 
    \multicolumn{1}{c|}{--}
\\ \hline
    \multicolumn{1}{|l|}{Sultana et al.~\cite{Sultana2021}} & 
    
    \multicolumn{1}{c|}{\checkmark} & 
    \multicolumn{1}{c|}{\checkmark} & 
    \multicolumn{1}{c|}{\checkmark} & 
    
    \multicolumn{1}{c|}{\cellcolor{gray!25}} & 
    \multicolumn{1}{c|}{--} & 
    \multicolumn{1}{c|}{\cellcolor{gray!25}} & 
    
    \multicolumn{1}{c|}{\checkmark} & 
    \multicolumn{1}{c|}{\checkmark} & 
    \multicolumn{1}{c|}{\checkmark}
\\ \hline
    \multicolumn{1}{|l|}{Santana et al.~\cite{santana2020smartbuildings}} & 
    
    \multicolumn{1}{c|}{\checkmark} & 
    \multicolumn{1}{c|}{\checkmark} & 
    \multicolumn{1}{c|}{\cellcolor{gray!25}} & 
    
    \multicolumn{1}{c|}{--} &
    \multicolumn{1}{c|}{--} & 
    \multicolumn{1}{c|}{\cellcolor{gray!25}} & 
    
    \multicolumn{1}{c|}{--} & 
    \multicolumn{1}{c|}{--} & 
    \multicolumn{1}{c|}{--}
\\ \hline
    \multicolumn{1}{|l|}{Alonso et al.~\cite{Alonso2020}} & 
    
    \multicolumn{1}{c|}{\checkmark} & 
    \multicolumn{1}{c|}{--} & 
    \multicolumn{1}{c|}{\cellcolor{gray!25}} & 
    
    \multicolumn{1}{c|}{--} & 
    \multicolumn{1}{c|}{--} & 
    \multicolumn{1}{c|}{\cellcolor{gray!25}} & 
    
    \multicolumn{1}{c|}{--} & 
    \multicolumn{1}{c|}{--} & 
    \multicolumn{1}{c|}{--}
\\ \hline
    \multicolumn{1}{|l|}{Sarabia-J\'acome et al.~\cite{Sarabia2020}} & 
    
    \multicolumn{1}{c|}{\checkmark} & 
    \multicolumn{1}{c|}{--} & 
    \multicolumn{1}{c|}{\cellcolor{gray!25}} & 
    
    \multicolumn{1}{c|}{--} & 
    \multicolumn{1}{c|}{--} & 
    \multicolumn{1}{c|}{\cellcolor{gray!25}} & 
    
    \multicolumn{1}{c|}{--} & 
    \multicolumn{1}{c|}{--} & 
    \multicolumn{1}{c|}{--}
\\ \hline
    \multicolumn{1}{|l|}{Xu et al.~\cite{Xu2018}} & 
    
    \multicolumn{1}{c|}{\checkmark} & 
    \multicolumn{1}{c|}{\checkmark} & 
    \multicolumn{1}{c|}{\cellcolor{gray!25}} & 
    
    \multicolumn{1}{c|}{\checkmark} & 
    \multicolumn{1}{c|}{--} & 
    \multicolumn{1}{c|}{\checkmark} & 
    
    \multicolumn{1}{c|}{\checkmark} & 
    \multicolumn{1}{c|}{\cellcolor{gray!25}} & 
    \multicolumn{1}{c|}{\cellcolor{gray!25}}
\\ \hline
    \multicolumn{1}{|l|}{Alves et al.~\cite{alves2020industry}} & 
    
    \multicolumn{1}{c|}{\checkmark} & 
    \multicolumn{1}{c|}{\checkmark} & 
    \multicolumn{1}{c|}{\checkmark} & 
    
    \multicolumn{1}{c|}{\cellcolor{gray!25}} & 
    \multicolumn{1}{c|}{\cellcolor{gray!25}} & 
    \multicolumn{1}{c|}{\cellcolor{gray!25}} & 
    
    \multicolumn{1}{c|}{--} & 
    \multicolumn{1}{c|}{--} & 
    \multicolumn{1}{c|}{--}
\\ \hline
    \multicolumn{1}{|l|}{Conroy et al.~\cite{conroy2022infection}} & 
    
    \multicolumn{1}{c|}{\checkmark} & 
    \multicolumn{1}{c|}{\checkmark} & 
    \multicolumn{1}{c|}{\checkmark} & 
    
    \multicolumn{1}{c|}{\cellcolor{gray!25}} &
    \multicolumn{1}{c|}{\cellcolor{gray!25}} & 
    \multicolumn{1}{c|}{\cellcolor{gray!25}} & 
    
    \multicolumn{1}{c|}{--} & 
    \multicolumn{1}{c|}{--} & 
    \multicolumn{1}{c|}{--}
\\ \hline
    \multicolumn{1}{|l|}{Bellocchio et al.~\cite{bellocchio2016smartseal}} & 
    
    \multicolumn{1}{c|}{\checkmark} & 
    \multicolumn{1}{c|}{\checkmark} & 
    \multicolumn{1}{c|}{\cellcolor{gray!25}} & 
    
    \multicolumn{1}{c|}{--} & 
    \multicolumn{1}{c|}{\cellcolor{gray!25}} & 
    \multicolumn{1}{c|}{\cellcolor{gray!25}} & 
    
    \multicolumn{1}{c|}{--} & 
    \multicolumn{1}{c|}{--} & 
    \multicolumn{1}{c|}{--}
\\ \hline
    \multicolumn{1}{|l|}{Shih et al.~\cite{shih2020warning}} & 
    
    \multicolumn{1}{c|}{\checkmark} & 
    \multicolumn{1}{c|}{\checkmark} & 
    \multicolumn{1}{c|}{--} & 
    
    \multicolumn{1}{c|}{\cellcolor{gray!25}} & 
    \multicolumn{1}{c|}{\cellcolor{gray!25}} & 
    \multicolumn{1}{c|}{\cellcolor{gray!25}} & 
    
    \multicolumn{1}{c|}{--} & 
    \multicolumn{1}{c|}{--} & 
    \multicolumn{1}{c|}{--}
\\ \hline
    \multicolumn{1}{|l|}{Gallagher et al.~\cite{gallagher2019intellimav}} & 
    
    \multicolumn{1}{c|}{\checkmark} & 
    \multicolumn{1}{c|}{\checkmark} & 
    \multicolumn{1}{c|}{--} & 
    
    \multicolumn{1}{c|}{\cellcolor{gray!25}} &
    \multicolumn{1}{c|}{\cellcolor{gray!25}} & 
    \multicolumn{1}{c|}{\cellcolor{gray!25}} & 
    
    \multicolumn{1}{c|}{--} & 
    \multicolumn{1}{c|}{--} & 
    \multicolumn{1}{c|}{--}
\\ \hline
    \multicolumn{1}{|l|}{Salhaoui et al.~\cite{Salhaoui2020}} & 
    
    \multicolumn{1}{c|}{\checkmark} & 
    \multicolumn{1}{c|}{\cellcolor{gray!25}} & 
    \multicolumn{1}{c|}{\cellcolor{gray!25}} & 
    
    \multicolumn{1}{c|}{--} & 
    \multicolumn{1}{c|}{--} & 
    \multicolumn{1}{c|}{\cellcolor{gray!25}} & 
    
    \multicolumn{1}{c|}{--} & 
    \multicolumn{1}{c|}{--} & 
    \multicolumn{1}{c|}{--}
\\ \hline
    \multicolumn{1}{|l|}{Shi et al.~\cite{Shi2019}} & 
    
    \multicolumn{1}{c|}{\checkmark} & 
    \multicolumn{1}{c|}{\cellcolor{gray!25}} & 
    \multicolumn{1}{c|}{\cellcolor{gray!25}} & 
    
    \multicolumn{1}{c|}{--} & 
    \multicolumn{1}{c|}{--} & 
    \multicolumn{1}{c|}{\cellcolor{gray!25}} & 
    
    \multicolumn{1}{c|}{--} & 
    \multicolumn{1}{c|}{--} & 
    \multicolumn{1}{c|}{--}
\\ \hline
    \multicolumn{1}{|l|}{Nguyen et al.~\cite{nguyen2021finger}} & 
    
    \multicolumn{1}{c|}{\checkmark} & 
    \multicolumn{1}{c|}{\checkmark} & 
    \multicolumn{1}{c|}{\checkmark} & 
    
    \multicolumn{1}{c|}{\cellcolor{gray!25}} &
    \multicolumn{1}{c|}{\cellcolor{gray!25}} & 
    \multicolumn{1}{c|}{\cellcolor{gray!25}} & 
    
    \multicolumn{1}{c|}{\cellcolor{gray!25}} & 
    \multicolumn{1}{c|}{\checkmark} & 
    \multicolumn{1}{c|}{\checkmark}
\\ \hline
    \multicolumn{1}{|l|}{Schumann et al.~\cite{schumann2012software}} & 
    
    \multicolumn{1}{c|}{\checkmark} & 
    \multicolumn{1}{c|}{\checkmark} & 
    \multicolumn{1}{c|}{\checkmark} & 
    
    \multicolumn{1}{c|}{\cellcolor{gray!25}} & 
    \multicolumn{1}{c|}{\cellcolor{gray!25}} & 
    \multicolumn{1}{c|}{\cellcolor{gray!25}} & 
    
    \multicolumn{1}{c|}{\cellcolor{gray!25}} & 
    \multicolumn{1}{c|}{\checkmark} & 
    \multicolumn{1}{c|}{\checkmark}
\\ \hline
    \multicolumn{1}{|l|}{Schubert et al.~\cite{schubert2021onorbit}} & 
    
    \multicolumn{1}{c|}{\checkmark} & 
    \multicolumn{1}{c|}{\checkmark} & 
    \multicolumn{1}{c|}{\checkmark} & 
    
    \multicolumn{1}{c|}{\cellcolor{gray!25}} & 
    \multicolumn{1}{c|}{\cellcolor{gray!25}} & 
    \multicolumn{1}{c|}{\cellcolor{gray!25}} & 
    
    \multicolumn{1}{c|}{\cellcolor{gray!25}} & 
    \multicolumn{1}{c|}{\checkmark} & 
    \multicolumn{1}{c|}{\checkmark}
\\ \hline
    \multicolumn{1}{|l|}{Agarwal et al.~\cite{agarwal2016making}} & 
    
    \multicolumn{1}{c|}{\checkmark} & 
    \multicolumn{1}{c|}{\checkmark} & 
    \multicolumn{1}{c|}{\checkmark} & 
    
    \multicolumn{1}{c|}{\cellcolor{gray!25}} &
    \multicolumn{1}{c|}{\cellcolor{gray!25}} & 
    \multicolumn{1}{c|}{\cellcolor{gray!25}} & 
    
    \multicolumn{1}{c|}{\cellcolor{gray!25}} & 
    \multicolumn{1}{c|}{--} & 
    \multicolumn{1}{c|}{--}
\\ \hline
    \multicolumn{1}{|l|}{Habibi et al.~\cite{habibi2019itelescope}} & 
    
    \multicolumn{1}{c|}{\checkmark} & 
    \multicolumn{1}{c|}{\checkmark} & 
    \multicolumn{1}{c|}{--} & 
    
    \multicolumn{1}{c|}{\cellcolor{gray!25}} & 
    \multicolumn{1}{c|}{\cellcolor{gray!25}} & 
    \multicolumn{1}{c|}{\cellcolor{gray!25}} & 
    
    \multicolumn{1}{c|}{\cellcolor{gray!25}} & 
    \multicolumn{1}{c|}{--} & 
    \multicolumn{1}{c|}{--}
\\ \hline
    \multicolumn{1}{|l|}{M\"{u}ller et al.~\cite{Muller2019}} & 
    
    \multicolumn{1}{c|}{\checkmark} & 
    \multicolumn{1}{c|}{\checkmark} & 
    \multicolumn{1}{c|}{--} & 
    
    \multicolumn{1}{c|}{\cellcolor{gray!25}} & 
    \multicolumn{1}{c|}{\cellcolor{gray!25}} & 
    \multicolumn{1}{c|}{\cellcolor{gray!25}} & 
    
    \multicolumn{1}{c|}{\cellcolor{gray!25}} & 
    \multicolumn{1}{c|}{--} & 
    \multicolumn{1}{c|}{--}
\\ \hline
    \multicolumn{1}{|l|}{Lu et al.~\cite{lu2020digitaltwin}} & 
    
    \multicolumn{1}{c|}{\checkmark} & 
    \multicolumn{1}{c|}{\checkmark} & 
    \multicolumn{1}{c|}{\cellcolor{gray!25}} & 
    
    \multicolumn{1}{c|}{\cellcolor{gray!25}} &
    \multicolumn{1}{c|}{\cellcolor{gray!25}} & 
    \multicolumn{1}{c|}{\cellcolor{gray!25}} & 
    
    \multicolumn{1}{c|}{--} & 
    \multicolumn{1}{c|}{--} & 
    \multicolumn{1}{c|}{--}
\\ \hline
    \multicolumn{1}{|l|}{Calancea et al.~\cite{calancea2019iassistme}} & 
    
    \multicolumn{1}{c|}{\checkmark} & 
    \multicolumn{1}{c|}{\cellcolor{gray!25}} & 
    \multicolumn{1}{c|}{\cellcolor{gray!25}} & 
    
    \multicolumn{1}{c|}{\cellcolor{gray!25}} &
    \multicolumn{1}{c|}{--} & 
    \multicolumn{1}{c|}{\cellcolor{gray!25}} & 
    
    \multicolumn{1}{c|}{--} & 
    \multicolumn{1}{c|}{--} & 
    \multicolumn{1}{c|}{--}
\\ \hline
    \multicolumn{1}{|l|}{Quintero et al.~\cite{Quintero2019}} & 
    
    \multicolumn{1}{c|}{\checkmark} & 
    \multicolumn{1}{c|}{\cellcolor{gray!25}} & 
    \multicolumn{1}{c|}{\cellcolor{gray!25}} & 
    
    \multicolumn{1}{c|}{\cellcolor{gray!25}} & 
    \multicolumn{1}{c|}{--} & 
    \multicolumn{1}{c|}{\cellcolor{gray!25}} & 
    
    \multicolumn{1}{c|}{--} & 
    \multicolumn{1}{c|}{--} & 
    \multicolumn{1}{c|}{--}
\\ \hline
    \multicolumn{1}{|l|}{Franklin et al.~\cite{franklin2014lida}} & 
    
    \multicolumn{1}{c|}{\checkmark} & 
    \multicolumn{1}{c|}{\cellcolor{gray!25}} & 
    \multicolumn{1}{c|}{\cellcolor{gray!25}} & 
    
    \multicolumn{1}{c|}{\cellcolor{gray!25}} & 
    \multicolumn{1}{c|}{\cellcolor{gray!25}} & 
    \multicolumn{1}{c|}{\cellcolor{gray!25}} & 
    
    \multicolumn{1}{c|}{\checkmark} & 
    \multicolumn{1}{c|}{\checkmark} & 
    \multicolumn{1}{c|}{\checkmark}
\\ \hline
    \multicolumn{1}{|l|}{Brumbaugh et al.~\cite{brumbaugh2019bighead}} & 
    
    \multicolumn{1}{c|}{\checkmark} & 
    \multicolumn{1}{c|}{--} & 
    \multicolumn{1}{c|}{--} & 
    
    \multicolumn{1}{c|}{\cellcolor{gray!25}} &
    \multicolumn{1}{c|}{--} & 
    \multicolumn{1}{c|}{\cellcolor{gray!25}} & 
    
    \multicolumn{1}{c|}{\cellcolor{gray!25}} & 
    \multicolumn{1}{c|}{\cellcolor{gray!25}} & 
    \multicolumn{1}{c|}{\cellcolor{gray!25}}
\\ \hline
    \multicolumn{1}{|l|}{Hegemier et al.~\cite{Hegemier2021}} & 
    
    \multicolumn{1}{c|}{\checkmark} & 
    \multicolumn{1}{c|}{\cellcolor{gray!25}} & 
    \multicolumn{1}{c|}{\cellcolor{gray!25}} & 
    
    \multicolumn{1}{c|}{--} & 
    \multicolumn{1}{c|}{--} & 
    \multicolumn{1}{c|}{\cellcolor{gray!25}} & 
    
    \multicolumn{1}{c|}{--} & 
    \multicolumn{1}{c|}{\cellcolor{gray!25}} & 
    \multicolumn{1}{c|}{\cellcolor{gray!25}}
\\ \hline
    \multicolumn{1}{|l|}{Gorkin et al.~\cite{gorkin2020sharkey}} & 
    
    \multicolumn{1}{c|}{\checkmark} & 
    \multicolumn{1}{c|}{\cellcolor{gray!25}} & 
    \multicolumn{1}{c|}{\cellcolor{gray!25}} & 
    
    \multicolumn{1}{c|}{\cellcolor{gray!25}} &
    \multicolumn{1}{c|}{\cellcolor{gray!25}} & 
    \multicolumn{1}{c|}{\cellcolor{gray!25}} & 
    
    \multicolumn{1}{c|}{--} & 
    \multicolumn{1}{c|}{--} & 
    \multicolumn{1}{c|}{--}
\\ \hline
    \multicolumn{1}{|l|}{Barachi et al.~\cite{barachi2020crowdsensing}} & 
    
    \multicolumn{1}{c|}{\checkmark} & 
    \multicolumn{1}{c|}{\cellcolor{gray!25}} & 
    \multicolumn{1}{c|}{\cellcolor{gray!25}} & 
    
    \multicolumn{1}{c|}{\cellcolor{gray!25}} &
    \multicolumn{1}{c|}{\cellcolor{gray!25}} & 
    \multicolumn{1}{c|}{\cellcolor{gray!25}} & 
    
    \multicolumn{1}{c|}{--} & 
    \multicolumn{1}{c|}{--} & 
    \multicolumn{1}{c|}{--}
\\ \hline
    \multicolumn{1}{|l|}{Niu et al.~\cite{niu2017adls}} & 
    
    \multicolumn{1}{c|}{\checkmark} & 
    \multicolumn{1}{c|}{\cellcolor{gray!25}} & 
    \multicolumn{1}{c|}{\cellcolor{gray!25}} & 
    
    \multicolumn{1}{c|}{\cellcolor{gray!25}} &
    \multicolumn{1}{c|}{\cellcolor{gray!25}} & 
    \multicolumn{1}{c|}{\cellcolor{gray!25}} & 
    
    \multicolumn{1}{c|}{--} & 
    \multicolumn{1}{c|}{--} & 
    \multicolumn{1}{c|}{--}
\\ \hline
    \multicolumn{1}{|l|}{Qiu et al.~\cite{qiu2020phm}} & 
    
    \multicolumn{1}{c|}{\checkmark} & 
    \multicolumn{1}{c|}{\cellcolor{gray!25}} & 
    \multicolumn{1}{c|}{\cellcolor{gray!25}} & 
    
    \multicolumn{1}{c|}{\cellcolor{gray!25}} &
    \multicolumn{1}{c|}{\cellcolor{gray!25}} & 
    \multicolumn{1}{c|}{\cellcolor{gray!25}} & 
    
    \multicolumn{1}{c|}{--} & 
    \multicolumn{1}{c|}{--} & 
    \multicolumn{1}{c|}{--}
\\ \hline
    \multicolumn{1}{|l|}{Cabanes et al.~\cite{cabanes2019autonomous}} & 
    
    \multicolumn{1}{c|}{\checkmark} & 
    \multicolumn{1}{c|}{\checkmark} & 
    \multicolumn{1}{c|}{\checkmark} & 
    
    \multicolumn{1}{c|}{\cellcolor{gray!25}} & 
    \multicolumn{1}{c|}{\cellcolor{gray!25}} & 
    \multicolumn{1}{c|}{\cellcolor{gray!25}} & 
    
    \multicolumn{1}{c|}{\cellcolor{gray!25}} & 
    \multicolumn{1}{c|}{\cellcolor{gray!25}} & 
    \multicolumn{1}{c|}{\cellcolor{gray!25}}
\\ \hline
    \multicolumn{1}{|l|}{Gao et al.~\cite{gao2016icn}} & 
    
    \multicolumn{1}{c|}{\checkmark} & 
    \multicolumn{1}{c|}{\checkmark} & 
    \multicolumn{1}{c|}{--} & 
    
    \multicolumn{1}{c|}{\cellcolor{gray!25}} & 
    \multicolumn{1}{c|}{\cellcolor{gray!25}} & 
    \multicolumn{1}{c|}{\cellcolor{gray!25}} & 
    
    \multicolumn{1}{c|}{\cellcolor{gray!25}} & 
    \multicolumn{1}{c|}{\cellcolor{gray!25}} & 
    \multicolumn{1}{c|}{\cellcolor{gray!25}}
\\ \hline
    \multicolumn{1}{|l|}{Amrollahi et al.~\cite{amrollahi2020aidex}} & 
    
    \multicolumn{1}{c|}{\checkmark} & 
    \multicolumn{1}{c|}{--} & 
    \multicolumn{1}{c|}{--} & 
    
    \multicolumn{1}{c|}{\cellcolor{gray!25}} &
    \multicolumn{1}{c|}{\cellcolor{gray!25}} & 
    \multicolumn{1}{c|}{\cellcolor{gray!25}} & 
    
    \multicolumn{1}{c|}{\cellcolor{gray!25}} & 
    \multicolumn{1}{c|}{\cellcolor{gray!25}} & 
    \multicolumn{1}{c|}{\cellcolor{gray!25}}
\\ \hline
    \multicolumn{1}{|l|}{Kemsaram et al.~\cite{kemsaram2020vision}} & 
    
    \multicolumn{1}{c|}{\checkmark} & 
    \multicolumn{1}{c|}{\checkmark} & 
    \multicolumn{1}{c|}{\cellcolor{gray!25}} & 
    
    \multicolumn{1}{c|}{\cellcolor{gray!25}} &
    \multicolumn{1}{c|}{--} & 
    \multicolumn{1}{c|}{\cellcolor{gray!25}} & 
    
    \multicolumn{1}{c|}{\cellcolor{gray!25}} & 
    \multicolumn{1}{c|}{\cellcolor{gray!25}} & 
    \multicolumn{1}{c|}{\cellcolor{gray!25}}
\\ \hline
    \multicolumn{1}{|l|}{Bayerl et al.~\cite{Bayerl2020}} & 
    
    \multicolumn{1}{c|}{\checkmark} & 
    \multicolumn{1}{c|}{--} & 
    \multicolumn{1}{c|}{\cellcolor{gray!25}} & 
    
    \multicolumn{1}{c|}{\cellcolor{gray!25}} & 
    \multicolumn{1}{c|}{--} & 
    \multicolumn{1}{c|}{\cellcolor{gray!25}} & 
    
    \multicolumn{1}{c|}{\cellcolor{gray!25}} & 
    \multicolumn{1}{c|}{\cellcolor{gray!25}} & 
    \multicolumn{1}{c|}{\cellcolor{gray!25}}
\\ \hline
    \multicolumn{1}{|l|}{Johny et al.~\cite{Johny2021}} & 
    
    \multicolumn{1}{c|}{\checkmark} & 
    \multicolumn{1}{c|}{\cellcolor{gray!25}} & 
    \multicolumn{1}{c|}{\cellcolor{gray!25}} & 
    
    \multicolumn{1}{c|}{\cellcolor{gray!25}} & 
    \multicolumn{1}{c|}{--} & 
    \multicolumn{1}{c|}{\cellcolor{gray!25}} & 
    
    \multicolumn{1}{c|}{--} & 
    \multicolumn{1}{c|}{\cellcolor{gray!25}} & 
    \multicolumn{1}{c|}{\cellcolor{gray!25}}
\\ \hline
    \multicolumn{1}{|l|}{Falcao et al.~\cite{falcao2021piwims}} & 
    
    \multicolumn{1}{c|}{\checkmark} & 
    \multicolumn{1}{c|}{\checkmark} & 
    \multicolumn{1}{c|}{\cellcolor{gray!25}} & 
    
    \multicolumn{1}{c|}{\cellcolor{gray!25}} & 
    \multicolumn{1}{c|}{\cellcolor{gray!25}} & 
    \multicolumn{1}{c|}{\cellcolor{gray!25}} & 
    
    \multicolumn{1}{c|}{\cellcolor{gray!25}} & 
    \multicolumn{1}{c|}{\cellcolor{gray!25}} & 
    \multicolumn{1}{c|}{\cellcolor{gray!25}}
\\ \hline
    \multicolumn{1}{|l|}{Hawes et al.~\cite{hawes2017strands}} & 
    
    \multicolumn{1}{c|}{\checkmark} & 
    \multicolumn{1}{c|}{\checkmark} & 
    \multicolumn{1}{c|}{\cellcolor{gray!25}} & 
    
    \multicolumn{1}{c|}{\cellcolor{gray!25}} &
    \multicolumn{1}{c|}{\cellcolor{gray!25}} & 
    \multicolumn{1}{c|}{\cellcolor{gray!25}} & 
    
    \multicolumn{1}{c|}{\cellcolor{gray!25}} & 
    \multicolumn{1}{c|}{\cellcolor{gray!25}} & 
    \multicolumn{1}{c|}{\cellcolor{gray!25}}
\\ \hline
    \multicolumn{1}{|l|}{Ali et al.~\cite{ali2016idviewer}} & 
    
    \multicolumn{1}{c|}{\checkmark} & 
    \multicolumn{1}{c|}{\cellcolor{gray!25}} & 
    \multicolumn{1}{c|}{\cellcolor{gray!25}} & 
    
    \multicolumn{1}{c|}{\cellcolor{gray!25}} &
    \multicolumn{1}{c|}{\cellcolor{gray!25}} & 
    \multicolumn{1}{c|}{\cellcolor{gray!25}} & 
    
    \multicolumn{1}{c|}{--} & 
    \multicolumn{1}{c|}{\cellcolor{gray!25}} & 
    \multicolumn{1}{c|}{\cellcolor{gray!25}}
\\ \hline
    \multicolumn{10}{c}{\checkmark = Adopted, -- = Partially adopted, \crule[gray!25]{1cm}{0.3cm} = Not adopted}
\end{tabular}
}
\end{table}

\hl{This section presents a survey of the selected ML-based systems. Our main goal is to understand why, how, and to what extent practitioners have adopted the DOA principles (Section~\ref{sec:doa-principles}).} The answer to these questions allows us to identify the systems' requirements that DOA principles satisfy and the practical approaches for implementing DOA, as well as to define the open challenges for building DOA-based systems (Section~\ref{sec:open-issues}). \hl{Table~\ref{tab:survey} shows the systems that adopt the DOA principles by contrasting them with the list of reviewed papers.} \hlr{The ML-based systems reported in these papers can fully, partially, or not adopt the DOA principles.} The level of adoption depends on the requirements these systems need to satisfy, the nature of the data they handle, and their deployment environments. This section details the adoption of each principle and sub-principle. \hl{Table~\ref{tab:survey} shows that nine out of the forty-five papers follow all or most DOA principles, reflecting an overall low adoption of DOA when implementing current ML-based systems. The lack of maturity and tooling around the DOA style, in contrast with popular SOA cloud-based deployments, is the reason behind this finding.} 

\subsection{Data as a First Class Citizen}
\label{subsec:datafirst}

\begin{figure}[t!]
	\centering
	\includegraphics[width=.7\textwidth]{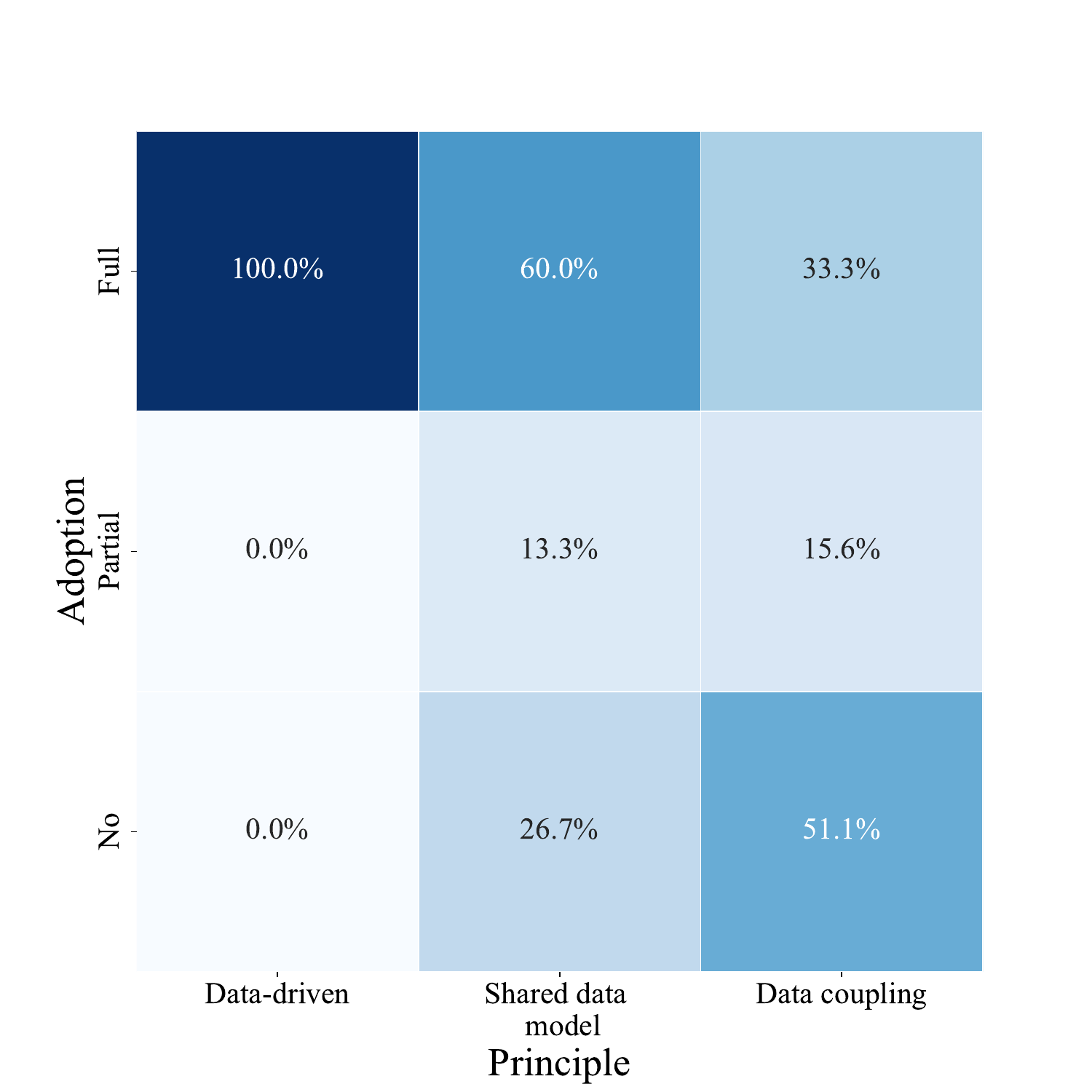}
        \caption{\hlr{Data as a first-class citizen principle adoption. All systems reviewed are data-driven, over 60\% at least partially adopt a shared data model, and just under half utilise data coupling.}}
	\label{fig:data-first}
\end{figure}

Figure~\ref{fig:data-first} shows the extent to which the reviewed papers adopt the \textbf{data as a first-class citizen} principle. We found that all reviewed papers report \textit{data-driven} systems, which is an expected result because ML-based systems are \textit{data-driven} by nature. \hl{All the surveyed systems contain one or multiple learning models because they want to satisfy data needs such as identifying anomalies, optimising operations, automating processes, etc. These systems' performance depends on the data flowing through them, and they accomplish their functional requirements using diverse data-driven methods.} For example, Schumann et al.~\cite{schumann2012software} use a Bayesian network for health monitoring of space vehicles (e.g., rovers), Sarabia-Jacome et al.~\cite{Sarabia2020} build a DL model for fall detection in Ambient Assisted Living (AAL) environments, Junchen et al.~\cite{Junchen2017} propose Pytheas as a data-driven approach that optimises the Quality of Experience (QoE) of applications based on network metrics, Agarwal et al.~\cite{agarwal2016making} present an RL-based decision-making system that Microsoft products use for different optimisation tasks, such as content recommendation system in MSN and virtual machine management in Azure Compute.

Our findings show that 60\% of the surveyed papers fully adopt the sub-principle of handling their information using \textit{shared data models}. \hl{These systems implement \textit{shared data models} using data streams or databases. Practitioners select data streams because these structures enable systems to manage and process information from continuous data sources.} That is the case of environments that require systems to process multimedia streams (e.g., video)~\cite{gao2016icn, Junchen2017,habibi2019itelescope,dai2019bigdl,falcao2021piwims}, sensors data~\cite{schumann2012software,cabanes2019autonomous,lu2020digitaltwin,shih2020warning,kemsaram2020vision,zhang202148learningadd,nguyen2021finger,herrero2022i40,conroy2022infection}, social media data~\cite{zhang2016emotion, Xu2018, Muller2019,shan2022poligraph}, and network metrics~\cite{santana2020smartbuildings, Sultana2021}. We observed that streams are common in systems built with dataflow architecture \cite{culler1986dataflow,paleyes2022fbpsoa}. A data flow architecture transforms data inputs using different components while they flow through the system. Herrero et al.~\cite{herrero2022i40} presents a good example of a system based on a dataflow architecture. They propose a platform for Industry 4.0 based on the RAI4.0 reference architecture~\cite{lopez2021datacentric} that models software and hardware components as stream producers and consumers. This platform handles data-intensive, continuous, and dynamic data sources from a water pump and its environment. Streams enable the system to manage and respond to real-time changes while providing predictive maintenance. Similarly, Sultana et al.~\cite{Sultana2021} present a software framework to detect Distributed Denial of Service (DDoS) attacks in real-time network traffic. This software relies on a Support Vector Machine (SVM) model deployed using the Acumos and ONAP platforms. The framework decouples components that exchange data between them via Kafka streams. \hl{System designers prefer databases as \textit{shared data models} when the system's functionalities require making decisions based on historical data records~\cite{bellocchio2016smartseal,agarwal2016making,hawes2017strands,lebofsky2019breakthrough,gallagher2019intellimav,alves2020industry,schubert2021onorbit}. The reason behind this design decision is that a shared data model implemented as a database facilitates the storage and management of historical records and provides an efficient communication medium between systems' components compared to synchronous API calls or Remote Procedure Calls (RPCs).} For example, Alves et al.~\cite{alves2020industry} propose a system for industrial predictive maintenance based on historical data collected from IoT devices. These devices produce and write monitoring data in a database. The system's components read such data for the respective maintenance decision-making. Schubert et al.~\cite{schubert2021onorbit} introduce a deep learning-based system used by the Texas Spacecraft Laboratory. The system's components store and retrieve data from AWS S3 buckets to generate large data sets of synthetic images that support on-orbit spacecraft operations. Hawes et al.~\cite{hawes2017strands} present a robotic platform that stores the Robotic Operative System (ROS) messages in a document-oriented database (i.e., MongoDB). These messages describe the status of the robot and its interactions with the environment. A monitoring component reads this history to predict future states. This robotic platform is the only reviewed work that uses the \textit{shared data model} and \textit{data coupling} principle for self-monitoring purposes.

Over thirteen per cent of the reviewed papers partially adopt a \textit{shared data model} between their systems' components. That is the case for systems combining heterogeneous storage technologies in shared data models~\cite{brumbaugh2019bighead,amrollahi2020aidex,karageorgou2020sentiment}. These systems must store, monitor, and trace the data status in different stages of its processing. \hl{Practitioners design and implement systems combining different storage technologies and models because the data structure and format change after each processing stage.} For example, the AIDEx platform~\cite{amrollahi2020aidex} queries an electronic medical records (EMR) database and passes such data to a set of microservices that predict the risk of patients' infection sepsis based on their data models. AIDEx creates patients' health data streams based on the prediction results stored in a MongoDB instance. The system uses this data for visualisation purposes. Karageorgou et al.~\cite{karageorgou2020sentiment} propose a system for multilingual sentiment analysis of Twitter streams. This platform uses RabbitMQ queues to collect tweets in different languages. The system then models such data as Kafka streams for sentiment analysis. Other systems that partially adopt a \textit{shared data model} are the ones that must collect data from spatially distributed sources~\cite{Sarabia2020, Bayerl2020, Alonso2020}. Systems distribute such data to satisfy low latency, resource constraints, and privacy requirements. Distributed nodes use the same models to store local data, but do not share this data between them. Subsection~\ref{subsec:decentralisation} reports more details on distributed and decentralised deployments.

Just over a quarter of the reviewed papers report systems whose components do not have shared data models. \hl{Practitioners design systems without shared data models because these systems are small and not data-intensive.} These are either implemented as monoliths in the cloud~\cite{niu2017adls,gorkin2020sharkey} or as edge architectures that use edge servers as gateways for data transmission and inference. Such edge solutions deploy trained learning models on embedded devices, like in the monitoring and prediction systems proposed by Quintero et al.~\cite{Quintero2019}, Salhaoui et al.~\cite{Salhaoui2020}, Hegemier et al.~\cite{Hegemier2021}, and Johny et al.~\cite{Johny2021}. Monolithic deployments follow a client-server pattern to collect sensor data and respond to requests. For example, Niu et al.~\cite{niu2017adls} deploy a recognition system of indoor daily activities for one-person household apartments. Some systems do not have shared data models but are complex and large and process big volumes of data. These rely on cloud data centres and follow well-known software architectures (e.g., microservices) to satisfy such requirements~\cite{franklin2014lida,ali2016idviewer, Shi2019,calancea2019iassistme,qiu2020phm,barachi2020crowdsensing}. Functionalities and data of systems’ subcomponents are encapsulated and hidden behind interfaces (e.g., service APIs). \hl{Designers prefer such implementations when they do not prioritise transparency and traceability requirements and when they rely on third-party cloud components for monitoring tasks (e.g., middleware solutions \cite{cabrera2017middleware}).}

Systems' components in 33\% of papers follow the \textit{data coupling} sub-principle. These systems' components interact by reading from and writing to data mediums. \hl{Engineers adopt this principle because systems must handle real-time data that needs to be processed continuously.} We found that components of real-time systems~\cite{schumann2012software, Junchen2017,cabanes2019autonomous,dai2019bigdl,karageorgou2020sentiment,nguyen2021finger, Sultana2021,herrero2022i40,brumbaugh2019bighead,conroy2022infection} act as subscribers and publishers of data to streams that represent the state of the data at different stages in a workflow. Stream-based systems make use of different technologies such as Apache Kafka~\cite{Junchen2017, Sultana2021,herrero2022i40} or Spark Streaming~\cite{Junchen2017,brumbaugh2019bighead}, sometimes adopting the underlying stream-based programming model for the entire system~\cite{dai2019bigdl}. Message queues (e.g., RabbitMQ) are also used in systems to collect heterogeneous data from different sources or enable interaction between components in parallel~\cite{karageorgou2020sentiment,nguyen2021finger}. Systems also adopt the \textit{data coupling} sub-principle to handle large amounts of data that are not possible to send through APIs or RPCs~\cite{lebofsky2019breakthrough,schubert2021onorbit} or when they analyse historical data~\cite{zhang2016emotion,agarwal2016making,alves2020industry,shan2022poligraph}. \hl{Practitioners design and implement data-coupled components using databases because systems and their components must process large pieces of data in batches}. The system proposed by Zhang et al.~\cite{zhang2016emotion} illustrates these types of \textit{data coupling} by combining streams and databases. Distributed Apache Kafka streams store social media data from Weibo and Chinese forums, which are processed using Apache Storm to enable real-time sentiment analysis. The system also offers batch processing where social media is stored using the Hadoop Distributed File System (HDFS) and an HBase database. The Apache Spark machine learning library (MLlib) performs distributed data analysis on top of this data. Systems that partially adopt \textit{data coupling} (i.e., 15.5\%)~\cite{gao2016icn,gallagher2019intellimav, Muller2019,habibi2019itelescope,amrollahi2020aidex,shih2020warning,brumbaugh2019bighead} are the ones where some components communicate through data mediums and others use traditional process calls. \hl{The reason behind such a hybrid approach is satisfying interoperability requirements while managing large and real-time data.} These systems use APIs or RPCs to communicate with distributed components or external entities, while the centralised system's components interact through data mediums. For example, systems can use API calls to collect data from heterogeneous data sources~\cite{Muller2019,amrollahi2020aidex} or to interact with end users~\cite{gao2016icn,gallagher2019intellimav}. Over half of the surveyed papers do not adopt \textit{data coupling}. These systems mostly use REST ~\cite{niu2017adls,Quintero2019,gorkin2020sharkey,santana2020smartbuildings,lu2020digitaltwin,barachi2020crowdsensing,kemsaram2020vision,Johny2021} or RPCs~\cite{franklin2014lida,bellocchio2016smartseal,hawes2017strands,Xu2018,Shi2019,calancea2019iassistme,Bayerl2020,qiu2020phm,Sarabia2020,Alonso2020,Salhaoui2020,falcao2021piwims,zhang202148learningadd,Hegemier2021} for communication. These systems' components hide the data behind their interfaces, which causes the data dichotomy issue impacting monitoring tasks~\cite{stopford2016data}.  

We found that not all systems where components \textit{share a data model} necessarily follow the \textit{data coupling} sub-principle. \hl{Different nodes in a distributed system can have the same data model even when they do not interact. Designers made this decision when systems must address low latency and data privacy requirements, but do not need to aggregate data from geographically distributed sources for training their models.} Some of these distributed systems create independent deployments that are in charge of the functionality of the whole system in a given geographical region or coverage area~\cite{bellocchio2016smartseal,hawes2017strands,lu2020digitaltwin,kemsaram2020vision}. Other distributed systems rely on edge devices that play the role of intermediate nodes that collect, preprocess, and transmit data to centralised servers as well as perform inference tasks when designers deploy cloud-trained learning models on the edge~\cite{Xu2018, Sarabia2020, Alonso2020, Bayerl2020,santana2020smartbuildings,zhang202148learningadd,falcao2021piwims,shan2022poligraph}.

\subsection{Prioritise Decentralisation}
\label{subsec:decentralisation}

\begin{figure}[t!]
	\centering
	\includegraphics[width=.7\textwidth]{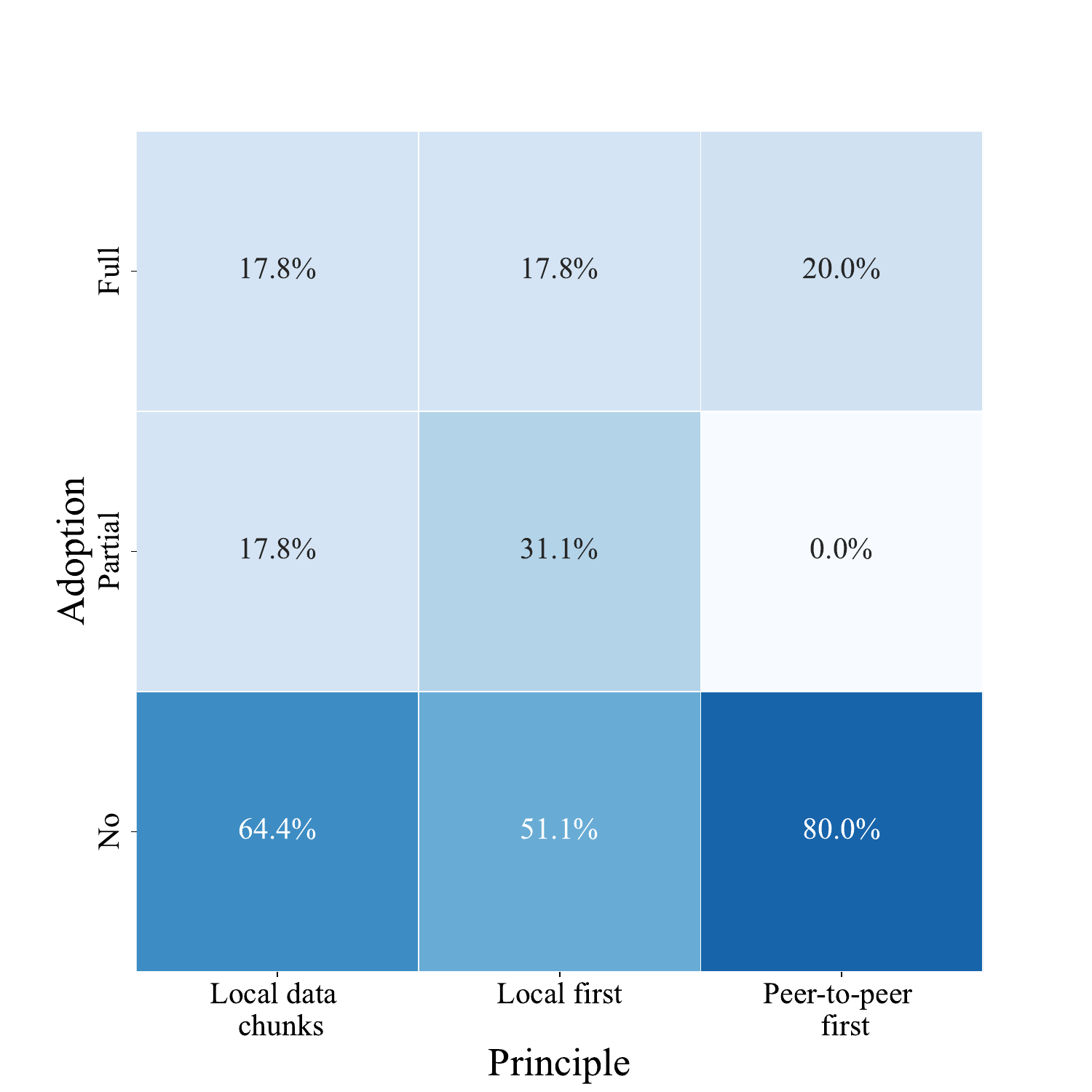}
        \caption{Adoption of the prioritising decentralisation principle. Most systems rely on centralised storage and processing to the detriment of the decentralisation sub-principles.}
	\label{fig:decentralised}
\end{figure}

Figure~\ref{fig:decentralised} presents the extent to which the papers \textbf{prioritise decentralisation} when deploying their ML-based systems. Almost eighteen per cent of the papers report systems where decentralised entities store \textit{local data chunks} with minimal presence of centralised entities. Decentralised approaches~\cite {zhang2016emotion, Xu2018,dai2019bigdl,lebofsky2019breakthrough,karageorgou2020sentiment,zhang202148learningadd,herrero2022i40,shan2022poligraph} rely on local data chunks, which provide partitioning and replication by design. \hl{Practitioners implement systems based on local data chunks to exploit such data partition and replication, enabling systems to manage computing resources efficiently and be fault-tolerant.} That is the case of the Breakthrough Listen program presented by Lebofsky~\cite{lebofsky2019breakthrough}. This program collects data from radio and optical telescopes to search for extraterrestrial intelligence (SETI). The search leverages different strategies, including ML models for modulation scheme classification and outlier detection. The paper reports the software and hardware architecture behind the data collection, reduction, archival, and public dissemination pipeline. This architecture consists of storage clusters and compute nodes that handle raw data volumes averaging 1PB daily. Decentralised storage nodes use Hierarchical Data Format version 5 (HDF5) to organise the large data sets in groups that facilitate data management along the pipeline. Another example of decentralisation is Poligraph~\cite{shan2022poligraph}, a system that detects fake news by combining ML models and human knowledge. Users send fake news detection requests to decentralised servers that run different ML models and collect news reviews from experts. These servers run the Byzantine Fault Tolerant (BFT) protocol to replicate users' requests and news to mitigate experts' unavailability (i.e., fault tolerance). \hl{Systems designers also use decentralisation to implement systems for processing intensive and sparse data sources~\cite{zhang2016emotion, Xu2018,karageorgou2020sentiment,herrero2022i40}.} For example, Zhang et al.~\cite{zhang2016emotion} propose a system for sentiment analysis of tweets from three Chinese cities (i.e., Beijing, Shanghai, Guangzhou and Chengdu). The decentralised storage and management of data leverages HBase and Kafka. \hl{We found that decentralised approaches must rely on distributed computing protocols and technologies to mitigate the challenges that arise from the absence of a central control entity. In addition to the technologies mentioned before, Karageorgou et al.~\cite{karageorgou2020sentiment} use Spark for scalable sentiment analysis on multilingual data from social media, Xu et al.~\cite{Xu2018} use Distributed Hash Tables (DHTs) to facilitate the retrieval of social data stored by topic for spam detection, and Herrero et al.~\cite{herrero2022i40} integrate Zookeeper, Kafka, and Apache Cassandra in a decentralised platform for a predictive maintenance industrial service.} 

Some systems partially decentralise the storage of data (17.8\% of the papers). These approaches exploit decentralisation properties (e.g., closeness to data owners) while having centralised control of the system~\cite{patel2014mobile,tabatabaee2022mecsurvey}. Partial decentralisation creates federated networks where collected data from sensor devices (e.g., IoT sensors, smartphones, wearables, etc.) is stored in local databases~\cite{bellocchio2016smartseal, Shi2019, Alonso2020, Hegemier2021} or encoded in local learning models trained using more powerful computing infrastructure~\cite{Junchen2017,santana2020smartbuildings, Salhaoui2020, Sarabia2020, Hegemier2021}. In both cases, edge servers preprocess and filter the collected data before transmitting it to back-end servers. Such central entities have a complete view of the system's state and perform more complex tasks (e.g., learning models training, job scheduling, resource allocation, etc.). \hl{System developers exploit these federated architectures to satisfy systems requirements such as data ownership, privacy, and security, as devices and edge nodes can decide when and how to transmit the information to back-end servers.} Santana et al.~\cite{santana2020smartbuildings} propose a crowd management system based on WiFi frames originating from people's smartphones in the context of the SmartSantander project~\cite{SANCHEZ2014217}. The system collects WiFi frames as streams preprocessed in edge nodes. This preprocessing includes data anonymisation and dimensionality reduction to protect people's identities and extract key features from the streams. A centralised processing tier stores and uses the preprocessed and anonymised data to train ML models (e.g., logistic regression, naive Bayesian, and random forest classifiers) for crowdsensing estimation. \hl{Practitioners also exploit local data storage to facilitate the deployment of ML-based solutions in resource-constrained environments and optimise cost.} For example, Alonso et al.~\cite{Alonso2020} implement a real-time monitoring edge system for smart farming, where farmers want to reduce storage and compute costs while monitoring the state of dairy cattle and feed grain. Edge servers in the system collect, buffer, and filter data from IoT devices to eliminate possible noise and discard duplicated frames. Such filtering also reduces transmission and storage costs.

Our findings show that most reviewed papers report systems that rely on centralised storage (i.e., 64.4\%). Some of these systems are deployed on single cloud servers~\cite{franklin2014lida,ali2016idviewer,agarwal2016making,gao2016icn,niu2017adls,calancea2019iassistme,habibi2019itelescope,brumbaugh2019bighead,Muller2019,gallagher2019intellimav,lu2020digitaltwin,barachi2020crowdsensing,gorkin2020sharkey,amrollahi2020aidex,qiu2020phm,shih2020warning,Sultana2021,schubert2021onorbit,conroy2022infection}, where all data is stored. \hl{Designers rely on centralised cloud servers because these are flexible and can handle Big Data requirements}. For example, Conroy et al.~\cite{conroy2022infection} present a system that monitors and predicts COVID-19 infections to aid military workforce readiness. This system monitored personnel of the US Department of Defence during the pandemic. It had around 10,000 users in ten months, collected 201 million hours of data, and delivered 599,174 total user-days of predictive service. Similarly, Calancea et al.~\cite{calancea2019iassistme} use the Google Cloud Datastore service to store the data behind iAssistMe, a platform to assist people with eye disabilities. Both systems assure scalability by increasing or reducing the number of servers based on the number of requests. Other systems are deployed on single resource-constraint devices like mobile phones~\cite{Quintero2019,Bayerl2020}, robots~\cite{schumann2012software,hawes2017strands,kemsaram2020vision}, or small processing units (e.g., Raspberry Pi)~\cite{cabanes2019autonomous,alves2020industry,nguyen2021finger,Johny2021,falcao2021piwims}. \hl{System designers implement this type of deployment to fit data problems that require online and fast inference and actuation, but where data storage is not needed or infeasible due to hardware or bandwidth constraints.} For example, Nguyen et al.~\cite{nguyen2021finger} present a neuroprosthetic hand with embedded deep learning-based control. The authors deploy a recurrent neural network (RNN) decoder on an NVIDIA Jetson Nano, enabling the implementation of the neuroprosthetic hand as a portable and self-contained unit with real-time control of individual finger movements. Data is not stored as it is processed online.

Above 17\% of systems prioritise \textit{local processing}. These systems deploy learning modes and other components on distributed computing nodes that process users' requests, make decisions, and provide systems' functionalities. \hl{The main reason for decentralised processing is to meet scalability and low latency requirements in data-intensive applications~\cite{zhang2016emotion, Junchen2017,lebofsky2019breakthrough,dai2019bigdl,karageorgou2020sentiment,zhang202148learningadd,shan2022poligraph,herrero2022i40}.} Dai et al.~\cite{dai2019bigdl} introduce BigDL, a distributed deep learning framework for big data platforms and workflows. Intel developed the BigDL framework to ease the application of deep learning in real-world data pipelines for its industrial users: Mastercard, World Bank, Cray, Talroo, UCSF, JD, UnionPay, Telefonica, GigaSpaces, and more. These companies handle dynamic and messy data that requires complex, iterative, and recurrent processing that is challenging to implement efficiently in centralised architectures. BigDL distributes training and inference on top of the scalable architecture of the Spark compute model. Spark partitions data across workers that form computing clusters. Workers apply map, filter, and reduce operations in a parallel fashion. The data partitioning and parallel processing enable the design and implementation of fault-tolerant and efficient systems based on learning models built from large data sets. \hlr{A few cases offer systems' functionalities based on the cooperation of distributed nodes.} These systems also follow the \textit{peer-to-peer first} sub-principle and correspond to the 20\% of the reviewed papers~\cite{zhang2016emotion, Junchen2017, Xu2018,dai2019bigdl,lebofsky2019breakthrough,karageorgou2020sentiment,zhang202148learningadd,shan2022poligraph,herrero2022i40}. \hl{System designers adopt \textit{local processing} and \textit{peer-to-peer first} principles to satisfy scalability, latency, fault-tolerance, and resource management requirements of systems that handle data from intensive and distributed sources.} That is the case with systems that analyse social media data, for example, sentiment analysis of tweets or spam detection~\cite{zhang2016emotion, Xu2018,karageorgou2020sentiment}, or the Breakthrough Listen program~\cite{lebofsky2019breakthrough} in the search for extraterrestrial intelligence (SETI). Adopting the \textit{peer-to-peer} sub-principle requires distributed computing technologies and protocols to handle the complexity of decentralised processing. Systems also use these technologies when they adopt the \textit{local first} sub-principle. Examples of the use of these technologies are Zhang et al.~\cite{zhang2016emotion}, Dai et al.~\cite{dai2019bigdl}, and Karageorgou et al.~\cite{karageorgou2020sentiment} using Spark, Junchen et al.~\cite{Junchen2017} based on Kafka, Xu et al.~\cite{Xu2018} using DHTs, Lebofsky et al.~\cite{lebofsky2019breakthrough} storing data on Hierarchical Data Format (HDF5), Shan et al.~\cite{shan2022poligraph} using the BFT protocol, and Herrero et al.~\cite{herrero2022i40} based on Zookeeper.

Over thirty-one per cent of the reviewed systems partially adopt the \textit{local first} sub-principle. Nodes in the network are in charge of different functionalities in these systems. Centralised servers are in charge of computationally expensive processing such as model training and updating~\cite{Xu2018, Quintero2019, Shi2019,kemsaram2020vision, Bayerl2020, Sarabia2020, Johny2021, Salhaoui2020, Hegemier2021}, global decision making~\cite{calancea2019iassistme, Alonso2020,santana2020smartbuildings}, and systems' orchestration~\cite{brumbaugh2019bighead, Sultana2021}. Centralised servers execute these tasks because they have enough computing power and a global view of the system. Central processing enables easier control of the components of the system, which can be challenging in fully decentralised systems as autonomous entities must self-govern. \hl{Practitioners opt for such central control to make systems robust as they rely on centralised servers to offer scalable and highly available solutions.} The works of Sultana et al.~\cite{Sultana2021} and Brumbaugh et al.~\cite{brumbaugh2019bighead} are good examples of central control. They encapsulate systems' functionalities in logically distributed containers that Kubernetes centrally controls. Kubernetes is the central platform that manages container instantiation, resource allocation, scheduling, and execution. Systems that partially adopt the \textit{local first} sub-principle use edge nodes or mobile devices to offer data collection and preprocessing functionalities. These nodes also offer inference capabilities when they host trained learning models. \hl{System developers use edge-computing architectures to enable local inference and fit low latency and data ownership requirements as learning models are closer to end users~\cite{tabatabaee2022mecsurvey,cabrera2023maaco}}. For example, Sarabia et al.~\cite{Sarabia2020} deploy an edge-based fall detection system for Ambient Assisted Living (AAL) environments. AAL systems process highly sensitive patient data and require very low processing time. The authors propose a 3-layer fog-cloud architecture composed of medical devices, fog nodes, and a cloud server. Deep learning models are deployed in fog nodes to detect patients' falls. The authors prove that the proposed platform performs better than a cloud baseline (i.e., better efficiency and response time). One or more fog nodes attend to each patient, creating a local area network to preprocess (e.g., filtering) the patient's data. Dedicated fog nodes enable data privacy and security by design. Kemsaram et al.~\cite{kemsaram2020vision} present the architecture design and development of an onboard stereo vision system for cooperative automated vehicles. The platform relies on a stereo camera that captures left and right images processed by pre-trained DNNs to perform object perception, lane perception, and free space perception. An object tracker then estimates different metrics over the classified images, such as depth and radial distance, relative velocity, and azimuth and elevation angle. The system implements a 4-layer architecture deployed in each vehicle to guide cooperative autonomous navigation.

\hlr{We found that over half of the papers describe centralised systems in which central nodes host all their components, which reflects the prevalent preference for the traditional client-server architecture. Some of these systems work in extreme resource-constrained scenarios that demand deployment on a single device~\cite{schumann2012software,hawes2017strands,cabanes2019autonomous,nguyen2021finger,falcao2021piwims}.} \hl{Engineers leverage such deployments in environments that require autonomous and self-contained systems because communication with back-end servers is limited or impossible.} That is the case of the work proposed by Schumann et al.~\cite{schumann2012software} that reports an ML-based system deployed in rovers to enable automatic vehicle monitoring in spatial missions. The Spirit and Opportunity Mars Exploration Rovers (MER) were autonomously functional on the surface of Mars for 6 and 14 years\footnote{NASA MER: \url{https://mars.nasa.gov/resources/spirit-and-opportunity-by-the-numbers/}}. These rovers performed data collection and navigation tasks. The STRANDS Core System~\cite{hawes2017strands} supports long-term autonomy (LTA) robot applications in security and care environments. \hlr{LTA refers to the capability of robots to operate continuously for multiple weeks.} STRANDS supported over 100 days of autonomous operations for robots performing navigation, human behaviour prediction, and activity recognition tasks. We also found that cloud servers host most centralised systems. \hl{Practitioners exploit cloud servers flexibility to address scalability and availability requirements~\cite{franklin2014lida,agarwal2016making,ali2016idviewer,bellocchio2016smartseal,gao2016icn,niu2017adls,gallagher2019intellimav, Muller2019,habibi2019itelescope,barachi2020crowdsensing,lu2020digitaltwin,amrollahi2020aidex,qiu2020phm,gorkin2020sharkey,shih2020warning,alves2020industry,schubert2021onorbit,conroy2022infection}.} That is the case of iTelescope~\cite{habibi2019itelescope}, which is an ML-based platform for real-time video classification. This platform collects data streams from the network traffic and processes them with two learning models for video identification and resolution classification. \hlr{iTeleScope worked in a campus network, served several hundred users, and demonstrated good performance with high concurrent streams.} Centralised entities rely on third-party entities to satisfy additional requirements such as monitoring and security. For example, AIDEx~\cite{amrollahi2020aidex} is a platform to predict patients’ risk of developing sepsis in the next 4 to 6 hours. AIDEx fetches patients’ records from a real-time EMR database and displays hourly sepsis risk scores for each patient. This platform is based on microservices for preprocessing data, executing the prediction learning model, and storing and visualising the prediction outcomes. The privacy and security of patients’ data are key system requirements, but designers leverage third-party mechanisms such as firewalls and virtual private clouds (VPCs).

Systems that follow the \textit{peer-to-peer first} sub-principle correspond to 20\% of the reviewed papers. The rest of the papers (i.e., 80\%) report systems that do not follow this sub-principle. Some of these are implemented as centralised architectures in the cloud~\cite{franklin2014lida,agarwal2016making,ali2016idviewer,gao2016icn, Muller2019,habibi2019itelescope,brumbaugh2019bighead,shih2020warning,barachi2020crowdsensing,amrollahi2020aidex,qiu2020phm, Sultana2021,schubert2021onorbit}. Central entities in these deployments orchestrate the interactions between systems' components. \hl{Designers opt for such centralised control as it enables robust and flexible systems that are easy to scale and highly available.} For example, Ali et al.~\cite{ali2016idviewer} present ID-Viewer, a surveillance system for identifying infectious diseases in Pakistan. Data collection, analysis, and visualisation components process data from different areas of the country. A central server hosts these components to enable spatio-temporal analysis of the causes and evolution of infectious diseases. \hl{Several works also describe systems as federated architectures~\cite{bellocchio2016smartseal,niu2017adls, Shi2019,calancea2019iassistme, Quintero2019,gallagher2019intellimav,gorkin2020sharkey,lu2020digitaltwin, Sarabia2020, Alonso2020, Salhaoui2020,alves2020industry,santana2020smartbuildings, Hegemier2021,conroy2022infection} without horizontal interactions between nodes at the same layer. Examples of these architectures are IoT deployments in different domains such as healthcare~\cite{niu2017adls, Quintero2019, Sarabia2020,conroy2022infection}, smart buildings~\cite{lu2020digitaltwin,santana2020smartbuildings}, farming~\cite{Alonso2020}, and predictive maintenance~\cite{alves2020industry}.} The architectures of these projects follow a layered pattern with vertical interaction between IoT devices, edge servers, and cloud servers. \hl{Designers use different layers to satisfy different system requirements}. Lower layers (i.e., IoT devices) are in charge of data collection, middle layers (i.e., edge servers) are in charge of data preprocessing, filtering, and transmission, and upper layers (i.e., cloud servers) are in charge of data aggregation, decision making, visualisation, and user interaction. There are also self-contained systems deployed in single constraint devices~\cite{schumann2012software,hawes2017strands,cabanes2019autonomous, Bayerl2020,kemsaram2020vision,nguyen2021finger,falcao2021piwims, Johny2021}. These are embedded systems that combine hardware and software components for specific functions in environments that require fast reactions (e.g., neuroprosthetic devices~\cite{nguyen2021finger}) or have limited connectivity with back-end servers (e.g., NASA's Mars rovers~\cite{schumann2012software}). 

\subsection{Openness}
\label{subsec:openness}

\begin{figure}[t!]
	\centering
	\includegraphics[width=.7\textwidth]{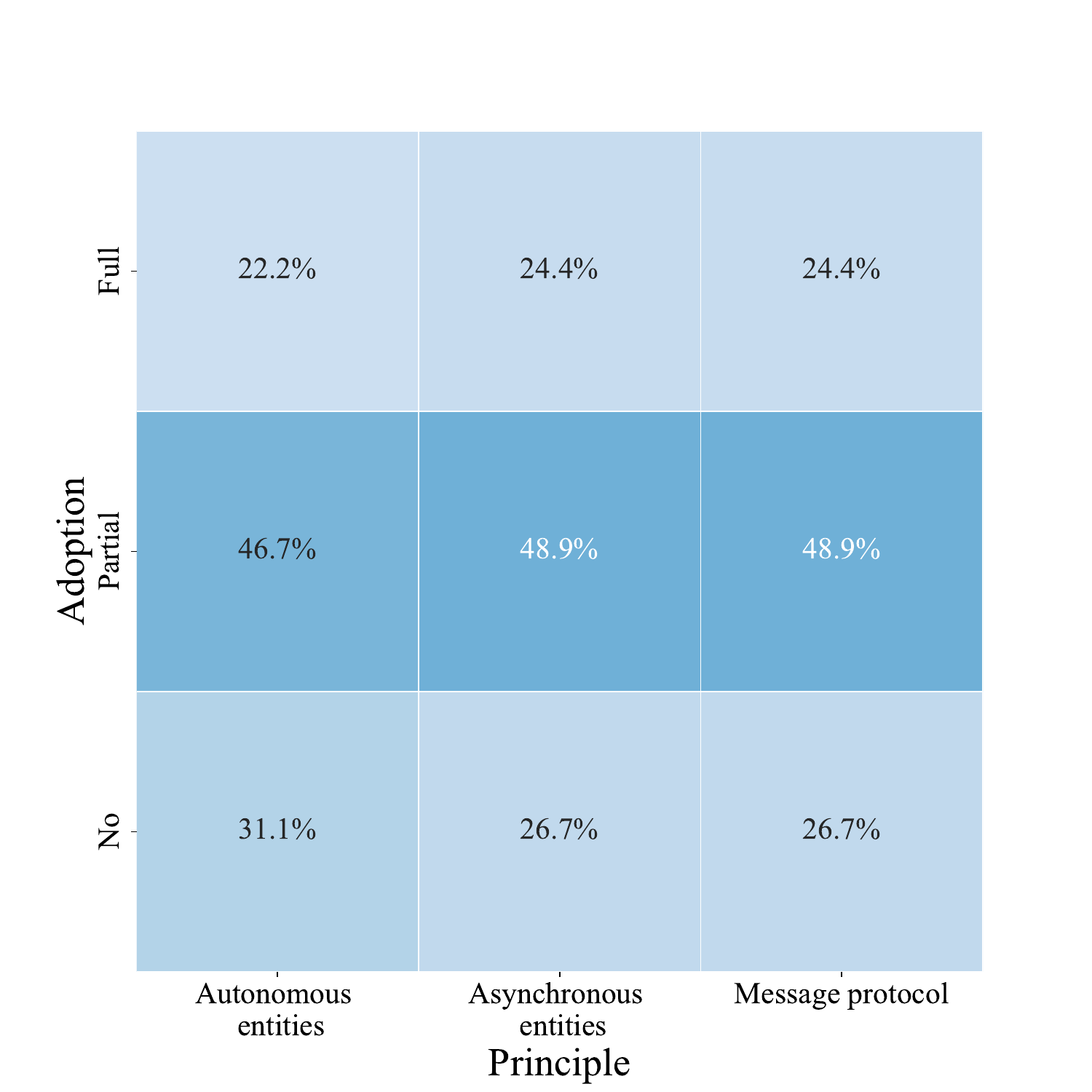}
        \caption{Adoption of openness principles. Most reviewed systems use autonomous entities, but only a third use them asynchronously. More than 50\% utilise message protocols.}
	\label{fig:openness}
\end{figure}

Figure~\ref{fig:openness} shows to what extent practitioners adopt the \textbf{openness} principle. Twenty-two per cent of the papers rely on flexible architectures where new components can join autonomously. These new components join and cooperate with the existing ones to achieve the system's goals. \hl{Practitioners exploit this flexibility when systems must be scalable and highly available. Dynamic requirements need on-demand extension of storage and computing capabilities~\cite{franklin2014lida,zhang2016emotion, Junchen2017, Xu2018,lebofsky2019breakthrough,dai2019bigdl,karageorgou2020sentiment, Sultana2021,herrero2022i40,shan2022poligraph}. Such flexibility causes security concerns as malicious entities can join the architecture. Designers must implement distributed protocols to address this risk.} For example, Shan et al.~\cite{shan2022poligraph} propose a decentralised architecture for Poligraph based on the BFT protocol, which adds servers and reviewers to the fake news detection process. \hlr{Such an adaptation responds to the request number variability, enabling scalability, while parallelism enables low latency.} \hl{A new node follows the BFT protocol to join the network. It sends messages to existing nodes to verify their identities, receives requests, and participates in the review consensus.} The Sifter platform~\cite{Xu2018} is an online spam detection system for social networks. It uses a recurrent neural network (RRN) for spam detection and Distributed Hash Tables (DHT) to address the fast-changing nature of topics and events. Social data is stored by decentralised servers and grouped by topics using the DHT structure to enable fast retrieval. Each server responds to spam detection requests using its local neural network. \hl{New servers can join by following the DHT protocol, which extends the storage capability of Sifter. A new node must contact an existing one, perform a lookup operation to find the closest nodes, and update these nodes' information and the network topology with its data.} Local spam detection fits low latency requirements as requests are solved closer to end users. We observe that the technologies and protocols used to handle decentralised storage and peer-to-peer cooperation (Section~\ref{subsec:decentralisation}) also support flexible architectures where entities can join autonomously. That is the case protocols such as DHT~\cite{Xu2018}, HDF5~\cite{lebofsky2019breakthrough}, Spark~\cite{dai2019bigdl}, or BFT~\cite{shan2022poligraph}, and architectural patterns such as dataflow~\cite{Sultana2021,Junchen2017}, the observer~\cite{franklin2014lida}, or publish/subscribe~\cite{zhang2016emotion,Junchen2017,karageorgou2020sentiment,Sultana2021,herrero2022i40}. 

There are systems based on federated~\cite{bellocchio2016smartseal,calancea2019iassistme,Shi2019,Alonso2020,santana2020smartbuildings,Sarabia2020,Salhaoui2020,zhang202148learningadd,Hegemier2021} and centralised~\cite{ali2016idviewer,niu2017adls,Quintero2019,gallagher2019intellimav,alves2020industry,gorkin2020sharkey,qiu2020phm,shih2020warning,lu2020digitaltwin,barachi2020crowdsensing,Johny2021,conroy2022infection} architectures that partially (i.e., 46\% of the papers) adopt the \textit{autonomous entities} sub-principle. \hl{These architectures have mechanisms to add new sensing devices or users at the lowest layer. These mechanisms consist of interoperable communication technologies that enable new actors to exchange messages with existing nodes. These technologies include WiFi, Bluetooth, Zigbee, 5G, etc.} However, these systems are not flexible to include more storage or compute nodes at the edge or cloud layers. \hl{Authors exploit such features to implement and deploy flexible systems where new sensors expand systems' data collection capabilities and new users expand systems' service coverage.} For example, Alonso et al.~\cite{Alonso2020} use Fiware\footnote{Fiware: \url{https://www.fiware.org/}} as a middleware to manage devices that transmit data using the ZigBee wireless standard. The collected data supports automated real-time decisions about the state of dairy cattle and the feed grain. Niu et al.~\cite{niu2017adls} developed the SensorBox device that integrates different IoT sensors to measure environmental variables. Supervised learning models are trained with such data to classify daily activities. Barachi et al.~\cite{barachi2020crowdsensing} deploys a crowdsensing platform based on DL to generate accident reports. Users are data consumers and collectors who can join the network by installing a mobile application. Calancea et al.~\cite{calancea2019iassistme} also require users to install an Android application to use the NLP-based platform that supports visually challenged people.

A third of the papers report ML-based systems whose components are not autonomous. \hlr{Practitioners avoid autonomous components when systems requirements do not demand dynamic changes on their components.} These systems are self-contained embedded systems~\cite{schumann2012software,hawes2017strands,cabanes2019autonomous,kemsaram2020vision, Bayerl2020,falcao2021piwims,nguyen2021finger} or rely on cloud-based architectures~\cite{agarwal2016making,gao2016icn,habibi2019itelescope, Muller2019,brumbaugh2019bighead,amrollahi2020aidex,schubert2021onorbit}. \hl{The authors deploy embedded systems to perform specific tasks with limited resources. A change in the tasks of these systems implies their redesign, including their hardware components.} That is the case of the robots of the STRANDS project~\cite{hawes2017strands}, specifically designed for navigation, human behaviour prediction, and activity recognition tasks. Similarly, Cabanes et al. ~\cite{cabanes2019autonomous}  particularly design an embedded system to detect parking events. It uses a single 3D sensor to collect the input data for an object recognition component. Cloud-based architectures rely on the cloud provider services to extend the storage or computing capabilities of the systems. This flexibility enables systems to satisfy availability and scalability requirements. For example, the Bighead framework~\cite{brumbaugh2019bighead} is based on Kubernetes to support the development of data-driven solutions across the whole Airbnb organisation. Designers achieve such flexibility at the cost of the cloud service providers' fees.

Almost a quarter of papers report systems that leverage \textit{asynchronous entities}~\cite{schumann2012software,franklin2014lida,zhang2016emotion, Junchen2017,dai2019bigdl,lebofsky2019breakthrough,karageorgou2020sentiment,schubert2021onorbit, Sultana2021,nguyen2021finger,herrero2022i40}. \hlr{Software components act as data producers and consumers when communicating asynchronously.} \hl{Developers adopt this sub-principle because it enables components not to block each other and operate independently. These do not wait for responses from other components but subscribe to relevant data producers. Systems based on these loosely coupled interactions satisfy low latency and big data processing requirements because asynchronous entities enable parallelisation and decentralised computing.} The papers that follow the \textit{asynchronous entities} sub-principle also follow the \textit{message exchange protocol} one. \hlr{Practitioners consider message exchange protocols because autonomous entities must communicate in open environments. Such protocols (e.g., MQTT, RabbitMQ, etc.) determine how entities exchange messages.} The sentiment analysis system proposed by Karageorgou et al.~\cite{karageorgou2020sentiment} relies on asynchronous components. Data from different intensive sources is collected using the publish/subscribe pattern, enabling different workers to process different data streams in parallel. This communication is based on RabbitMQ and Apache Kafka standards, while Spark enables a configurable number of workers. Similarly, BigDL~\cite{dai2019bigdl} support large-scale distributed training of deep learning models. Worker nodes constitute a Spark cluster to compute and aggregate local gradients for each model in parallel. Spark provides a data-parallel functional computing model that defines how workers operate. Workers perform map, reduce, and filter operations to transform data represented as Resilient Distributed Datasets (RDDs). Robotic systems are also based on asynchronous components because robots perform tasks in parallel. For example, a rover must move while monitoring its sensors' status~\cite{schumann2012software}. Similarly, the neuroprosthetic hand proposed by Nguyen et al.~\cite{nguyen2021finger} consists of three parallel threads that perform data acquisition, preprocessing, and motor decoding tasks. \hlr{A queue-based protocol drives the communication between these systems' components.} Data coupling (Section~\ref{subsec:datafirst}) enables asynchronous interactions by design as components write data to and read data from data mediums. We observe a high correlation between the \textit{asynchronous entities} and the \textit{data coupling} subprinciples, as most papers that follow the former also follow the latter.

Half of the papers report systems that partially adopt the \textit{asynchronous entities} and the \textit{message exchange protocol} sub-principles. \hl{Designers implement asynchronous communication because systems must listen to requests continuously or notify users at any time~\cite{calancea2019iassistme,barachi2020crowdsensing,gorkin2020sharkey,shih2020warning,shan2022poligraph}.} iAssistMe~\cite{calancea2019iassistme} relies on the pub/sub communication pattern to interact with visually challenged people and support their daily activities. The system uses RabbitMQ and the Advanced Message Queuing Protocol (AMQP) to handle data producers and consumers. Sharkeye~\cite{gorkin2020sharkey} uses the AWS Simple Notification Service (SNS) to send messages warning people about the presence of sharks via smartwatches. Users subscribe to the notification service following the pub/sub pattern. \hl{Other authors implement systems that use asynchronous entities to listen to data sources (e.g., sensor devices) that can produce data at any time~\cite{agarwal2016making,bellocchio2016smartseal,niu2017adls, Shi2019, Quintero2019, Muller2019,gallagher2019intellimav,habibi2019itelescope,lu2020digitaltwin,alves2020industry,santana2020smartbuildings, Salhaoui2020, Sarabia2020,qiu2020phm, Alonso2020,zhang202148learningadd,conroy2022infection}.} Bellocchio et al.~\cite{bellocchio2016smartseal} present the SEAL project to provide home automation based on ML for building energy management and safety. SmartSEAL is the platform that manages IoT devices deployed in the buildings. It offers asynchronous and synchronous communication using pub/sub and client-server protocols. The Robot Operative System (ROS) supports the pub/sub mechanism where sensor nodes write to and read from a topic. They extend ROS with a central entity (i.e., roscore), which provides device naming and registration services. Synchronous API calls enable the consumption of these services. Alves et al.~\cite{alves2020industry} propose an industrial system based on IoT nodes capable of acquiring and transmitting machine status parameters using the pub/sub architectural pattern (i.e., the MQTT protocol). IoT nodes post their readings in a shared database, which the system's components read asynchronously. Gallagher~\cite{gallagher2019intellimav} uses synchronous and asynchronous mechanisms in the IntelliMaV platform. IntelliMav is a cloud-based system that applies machine learning to verify the performance of energy conservation measures in near real-time. The system is deployed in the cloud and follows a 3-layer architecture. The user tier is the access point for users through a web browser that presents an interface for model training and deployment. The cloud tier is a virtual private cloud and hosts the application infrastructure. This tier processes the application requests by running R code exposed through APIs. The site tier represents the pipelines that collect data from the industrial site. These pipelines push the collected data to the cloud storage asynchronously for further processing. The smart diary tracer platform~\cite{Alonso2020} uses different protocols to collect data from diverse sources. These protocols include Zigbee, LoRa, WIFI, Bluetooth, and 3G, which support communication with IoT sensors deployed on the crops and the barns used to feed the livestock, the cattle to monitor their health, the factories for the traceability of packaged products and energy monitoring, and the trucks that transport the dairy products. Sarabia et al.~\cite{Sarabia2020} propose a 3-layer fog-cloud architecture where new healthcare devices join using low-power wireless technologies such as Bluetooth, Zigbee, 6LoWPAN, and Wi-Fi. The Bluetooth protocol is also used by Quintero et al.~\cite{Quintero2019} to include new devices to collect photoplethysmography data and heart rate monitoring.

The rest of the reviewed papers (i.e., 26.7\%) report synchronous systems. These rely on traditional client/server architectures and communication patterns such as RPC or API calls. \hl{The authors opt for this design decision when systems are small, self-contained, and attend synchronous requests~\cite{hawes2017strands,cabanes2019autonomous, Bayerl2020,kemsaram2020vision, Johny2021, Hegemier2021,falcao2021piwims}.} Hegemier et al.~\cite{Hegemier2021} propose a real-time danger avoidance systems for robots. They deploy a neural network in an edge server, which receives prediction requests from robots in HTTP format. Requests include an image of the robot environment and its status information (e.g., damage report). The edge server responds synchronously with a classification of the image. Johny et al.~\cite{Johny2021} propose a system for metastasis detection. They deploy a deep learning model in a Raspberry Pi, which performs image classification tasks. Users send their requests (i.e., images) to an embedded web server installed in the Pi. Components of larger systems also interact synchronously. These are usually deployed in the cloud and follow microservices architectures. They adopt layered and modular architectures that satisfy scalability and availability requirements~\cite{ali2016idviewer,gao2016icn,brumbaugh2019bighead,amrollahi2020aidex, Xu2018}. For example, the AIDEX~\cite{amrollahi2020aidex} system adopts an architecture to predict patients’ risk of developing sepsis. The system encapsulates each functionality as a microservice. The system exposes microservices for preprocessing data, executing the prediction algorithm, storing the prediction outcomes, and visualising outcomes. The system orchestrates these components using API calls to offer end-to-end capabilities such as sepsis prediction.

\subsection{Summary}

\begin{table}[]
\caption{\hl{Summary of our paper findings against the research questions we formulate in Section~\ref{sec:survey-method} for each DOA principle. The goal of this survey is to determine \textit{to what extent and how researchers have adopted the DOA principles to implement current ML-based systems, and what are the motivations behind such adoption?}}}
\label{tab:summary}
\resizebox{\textwidth}{!}{%
\begin{tabular}{|l|l|l|l|}
\hline
\multicolumn{1}{|c|}{\textbf{\begin{tabular}[c]{@{}c@{}}DOA \\ Principle\end{tabular}}}  & 
\multicolumn{1}{c|}{\textbf{\begin{tabular}[c]{@{}c@{}}Adoption Extent\\ (RQs To what extent?) \\ (See Figures~\ref{fig:data-first},~\ref{fig:decentralised}, and~\ref{fig:openness} and Table~\ref{tab:survey} ) \end{tabular}}} & 
\multicolumn{1}{c|}{\textbf{\begin{tabular}[c]{@{}c@{}}Adoption Motivation\\ (RQs Why?)\end{tabular}}} & 
\multicolumn{1}{c|}{\textbf{\begin{tabular}[c]{@{}c@{}}Adoption in Practice\\ (RQs How?)\end{tabular}}} \\ 
\hline
\begin{tabular}[c]{@{}l@{}}Data as a First\\ Class Citizen\end{tabular}  & 
\begin{tabular}[c]{@{}l@{}}
- All reviewed systems are \textit{data-driven.}\\
- Almost 75\% of papers fully or partially adopt the\\ \textit{shared data model} sub-principle.\\ 
- Less than half of the systems adopt \textit{data coupling.}\\
\end{tabular} & 
\begin{tabular}[c]{@{}l@{}}
ML-based systems use data-driven algorithms \\ to satisfy their functional requirements. These \\ systems must handle continuous data sources \\ and historical data, usually large and \\ dynamic. Data coupling fits better these data \\ management requirements.
\end{tabular} & 
\begin{tabular}[c]{@{}l@{}}
Practitioners use diverse data-driven algorithms \\ to provide different functionalities. For example, \\ DL models for fall detection, RL models \\ for resource optimisation, Bayesian networks \\ for rovers monitoring, etc. Designers implement \\ \textit{shared data models} and \textit{data coupling} \\ using databases (e.g., HDF5, HBase, etc.),\\ streams (e.g., Apache Kafka, Spark Streaming,\\ etc.), and message queues (e.g., RabbitMQ).
\end{tabular} \\ 
\hline
\begin{tabular}[c]{@{}l@{}}Prioritise \\ Decentralisation\end{tabular} & 
\begin{tabular}[c]{@{}l@{}}
- Only 35\% of works fully or partially adopt the\\ \textit{local data chunks} sub-principle.\\
- Almost half of the reviewed papers fully\\or partially perform \textit{local first} processing.\\
- Only 20\% of systems rely on components that\\prioritise \textit{peer-to-peer} collaboration.\\
\end{tabular} & 
\begin{tabular}[c]{@{}l@{}}
Practitioners implement decentralised or \\ federated architectures because these enable \\ systems to manage resources efficiently and be \\ fault-tolerant thanks to data partitioning and \\ replication. In addition, \textit{local data chunks} \\ improve systems privacy and security as the \\ data ownership does not change.
\end{tabular} & 
\begin{tabular}[c]{@{}l@{}}
Practitioners rely on distributed storage and \\ computing technologies and protocols to \\ implement and manage decentralised or \\ federated architectures. Examples of these \\ technologies and protocols are Apache Kafka, \\ HDFS, DHT, and BFT.
\end{tabular} \\ 
\hline
\begin{tabular}[c]{@{}l@{}}Openness\end{tabular} & 
\begin{tabular}[c]{@{}l@{}}
- Almost 70\% of the reviewed systems rely on\\ \textit{autonomous entities}.\\
- Almost 75\% or papers rely on \textit{asynchoronous}\\ \textit{entities} that communicate using a \textit{message protocol}.\\
- Hybrid architectures prevail, combining static\\and autonomous entities based on synchronous and\\asynchronous communication protocols.\\
\end{tabular} & 
\begin{tabular}[c]{@{}l@{}}
Dynamic environments like the ones we find \\ in the real world require flexible architectures \\ where new components (e.g., IoT sensors) can \\ join at runtime. Practitioners adopt the \textit{openness} \\ principle to design and implement such \\ architectures. \textit{Autonomous entities} usually \\ communicate in an \textit{asynchronous} fashion using \\ different \textit{message exchange protocols.}
\end{tabular} & 
\begin{tabular}[c]{@{}l@{}}
Practitioners use distributed computing (e.g., \\ Apache Kafka, DHT, etc.) and communication \\ technologies to implement \textit{open} systems. \\ Examples of communication technologies are \\ Wi-Fi, Bluetooth, Zigbee, and 5G. Protocols \\ like  MQTT and RabbitMQ, work on top of these \\ technologies to enable \textit{asynchronous} \\ \textit{communication} between \textit{autonomous entities}.
\end{tabular} \\ 
\hline
\end{tabular}
}
\end{table}

Table~\ref{tab:summary} summarises our survey findings against the research question we formulated in Section~\ref{sec:survey-method}. This section highlights the key findings and provides practical advice towards implementing and deploying Data-Oriented ML-based systems in the real world. \hl{We found that only one-fifth of the reviewed ML-based systems follow all or most DOA principles~\cite{zhang2016emotion, Junchen2017,lebofsky2019breakthrough,dai2019bigdl,karageorgou2020sentiment,zhang202148learningadd,shan2022poligraph,herrero2022i40}, which reflects a low adoption of DOA principles when implementing current ML-based systems. Despite this low adoption, we found that designers exploit DOA principles in the reviewed systems to satisfy requirements that are becoming increasingly common for data-driven systems. These systems address data requirements that demand managing big data from distributed and intensive sources, low-latency processing tasks, and efficient management of storage and computing resources.} We observed that systems based on architectural patterns such as dataflow and publish/subscribe are more data-oriented and follow more DOA principles. These system designs focus on data exchange between components (i.e., data first), which does not assume any coupling on the control flow level. \textbf{Practical advice:} Developers focus first on the data while creating data-intensive systems, and operations are secondary. Dataflow and publish/subscribe patterns are well-suited for designing and implementing DOA systems.

\hl{We found that the \textbf{data as a first-class citizen} is the most adopted principle among the reviewed papers, mainly because of the type of requirements ML-based systems have. These systems are \textit{data-driven} by nature, implying that they must manage large and dynamic data sets. \textit{Shared data models} and {data coupling} sub-principles fit better this type of requirements}. Engineers adopt the \textbf{data as a first-class citizen} principle because it enables efficient interactions between system components. This design decision avoids data transmission between components as payloads of direct calls (e.g., API calls). Systems' components act as producers and consumers of data stored in \textit{shared data models}. Such \textit{data coupling} offers asynchronous interactions by design, which makes components autonomous and non-blocking for each other. \textit{Data coupling} also addresses low latency requirements and resource constraints because autonomous components can process large data sets in parallel. The nature of the data systems handled influences the selection of the \textit{shared data models} that the systems' components nurture. \textit{Data coupling} based on databases is appropriate for systems that work with historical data, while streams fit better systems that handle continuous data from dynamic sources. An engineer can design a system with one or more shared data models. This decision depends on how the system processes the data from inputs to outputs. \textbf{Practical advice:} \textit{Data coupling} enables systems to address big data processing requirements in \textit{data-driven} systems. Databases, streams, and message queues are examples of data mediums to consider when designing data-first systems. These data mediums play the role of \textit{shared data models}, which practitioners implement using technologies such as Apache Kafka, Spark Streaming, HDF5, RabbitMQ, or HBase.

\hl{Our results show that the \textbf{prioritise decentralisation} is less popular than the \textbf{data as a first-class citizen} one because centralised cloud deployments prevail as the preferred choice when deploying systems nowadays. Practitioners have this strong preference because cloud platforms are flexible and mature enough to automate cumbersome infrastructure management processes. Decentralised architectures (e.g., edge computing) require communities to develop tools to facilitate their management for practitioners.} We found that centralised architectures are the preferred design choice when deploying ML-based systems to the detriment of the \textbf{prioritise decentralisation} principle. Cloud platforms cope well with the most relevant requirements that ML-based systems have nowadays as they provide flexibility, high availability, scalability, and, most of the time, satisfy current low-latency requirements. However, it is natural to expect these requirements to be more critical and severe as data becomes more available, users require lower latency, and applications become more complex (e.g., digital twins, augmented reality, etc.). Further research on decentralised architectures is necessary to enable the benefits of decentralisation while making it a feasible option for deployment. Reviewed works relying on edge computing already leverage federated architectures with increasingly powerful edge servers that offer \textit{local data storage and processing}. However, the \textit{peer-to-peer} interaction between nodes is missing in most cases. Cloud servers are the backend and still play the central storage and processing roles. A collaboration between nodes at lower layers in edge architectures can enable more sustainable systems by exploiting the storage and computing power of everyday devices~\cite{Kleppmann2019localfirst}. \textbf{Practical advice:} The absence of a central orchestrator in favour of direct communication between any two nodes of the system is a straightforward way to move to decentralisation. The authors of the reviewed papers used distributed storage technologies (e.g., Apache Kafka, HDFS, etc.) and distributed protocols (e.g., DHTs and BFT) to implement decentralised solutions.

Systems that collect data from distributed devices (e.g., sensors) are usually \textbf{open}. These systems use communication technologies such as Wi-Fi, Bluetooth, Zigbee, and 5G, among others, to add new sensors and data sources. Architectures are more closed and static at upper levels, whereas software components are static entities that rely on synchronous communication via RPC or API calls. Data coupling and open environments have a strong correlation. Reading from and writing to data mediums is an \textit{asynchronous communication} process where systems' components operate as data consumers and producers. This process requires components to utilise \textit{message exchange protocols} that describe how and when to read and write data. Such protocols also enable seamless and flexible architectures where components can join or leave autonomously. \textbf{Practical advice:} The use of message exchange protocols and data coupling results in systems that are open and flexible by design. It enables data availability and horizontal scalability as systems add resources on demand. Communication protocols such as MQTT and RabbitMQ are well-known tools for building open systems.

\subsection{Threats to validity}

In this section, we discuss threats to the validity and limitations of our work.

\textbf{Internal validity.} We used a multi-stage selection process to find papers for the survey. We have followed an established methodology (SLR, \cite{kitchenham2007guidelines, Kitchenham20132049}). However, this approach can have drawbacks.

There might be missed terms in the search queries that we have run against digital libraries. To mitigate this risk, we have iteratively improved our search procedure multiple times based on the analysis of the search output.

The search functionalities between different databases used in our automatic search differ. We addressed this risk by including as many publicly available library APIs as possible, even if their content overlaps. Our tool is publicly available and easy to extend to include new sources in the future to ensure quality and transparency.

\hl{The number of retrieved works required to automate the filtering process. We used the state-of-the-art algorithm Lbl2Vec to implement the automatic filter. However, this semantic filter could exclude relevant works. We tested the Lbl2Vec algorithm under different configurations to determine the best possible result. The Lbl2Vec algorithm requires defining classes for categorising papers as included and excluded. We explore different configurations for the classes and validate the outputs against the desired research works we expected to include. Furthermore, we used the snowballing step in the methodology to identify relevant papers we could have missed.}

\textbf{External validity.} We can generalise the conclusions in our study beyond the projects described in the reviewed papers and serve as general advice for ML engineers. Here, we discuss potential threats to such generalizability and how we addressed them.

The scientific literature does not describe many ML deployments. Some deployments are presented as blog posts, while others are not published anywhere. We did not include such projects in our study and dismissed some published reports because they omitted information about their software architecture from the paper. Nevertheless, our survey covers a significant subject of areas, including ML deployments across fields ranging from healthcare to autonomous driving (see Table \ref{tab:synonyms}). We have also taken great care not to impose unjustified limits on the search procedure to improve coverage.

The analysis and conclusions of this paper can be prone to personal biases as it relies on the authors' expertise. Following the advice by Kitchenham et al. \cite{kitchenham2009systematic}, the authors of this survey reviewed each other's study selection, analysis and conclusions. To further alleviate this risk, we sought feedback on our work, presenting intermediate results internally to our research group and at external scientific events.

\section{Open Research Challenges}
\label{sec:open-issues}

In the previous section, we observed that while the DOA principles offer the desired properties that enable ML-based systems to achieve demanding requirements at deployment, their adoption rates are mostly low. More research efforts are needed to advance the community's understanding of DOA, its strengths and weaknesses, why and how to build DOA systems, and the advantages of the capabilities DOA offers. Table~\ref{tab:open-challenges} summarises the identified DOA limitations from our survey. The rest of this section discusses them and formulates future research directions accordingly.

\begin{table}[]
\caption{Survey findings, open challenges, and possible actions to address them.}
\label{tab:open-challenges}
\resizebox{\textwidth}{!}{%
\begin{tabular}{|l|l|l|}
\hline
\multicolumn{1}{|c|}{\textbf{Survey Findings}} & 
\multicolumn{1}{c|}{\textbf{Open Challenge}} & 
\multicolumn{1}{c|}{\textbf{Research Directions}} \\ 
\hline
\begin{tabular}[c]{@{}l@{}}
A few papers fully adopt the DOA \\ principles. \textbf{(RQ1, RQ2, RQ3)}
\end{tabular} & 
\begin{tabular}[c]{@{}l@{}}
The DOA style is not fully \\ integrated with other ML areas \\ and the practitioners' workflow.
\end{tabular} & 
\begin{tabular}[c]{@{}l@{}}
The DOA style offers a range of desirable properties for multiple \\ software systems. Integrating the DOA style within other AI and ML \\ areas is an interesting and open research area. Alternative research \\ paths in this area are:\\ \\
- Develop DOA-based systems as enablers of ML pipelines (e.g., emulators, \\ Bayesian optimisation, etc.).\\ 
- Explore the intersection between decentralised DOA-based systems \\ and fields like edge computing, federated learning, self-adaptive systems, \\ and causal inference.\\ 
- Case studies to define the role of DOA-based systems in data management \\ and governance.\\ 
- Study how DOA principles can support LLMs in collecting data for \\ finetuning or enriching prompting.
\end{tabular} \\ 
\hline
\begin{tabular}[c]{@{}l@{}}
Only one of the reviewed papers \\ uses a \textit{shared data model} and \textit{data} \\ \textit{coupling} with monitoring purposes. \\ \textbf{(RQ1)}
\end{tabular}  & 
\begin{tabular}[c]{@{}l@{}}
Current systems do not exploit \\ the DOA style monitoring \\ properties.
\end{tabular} & 
\begin{tabular}[c]{@{}l@{}}
One of the key features that \textit{shared data models} and \textit{data coupling} \\ enable is the capability of systems self-monitoring. Research paths in \\ this direction are:\\ \\ 
- Develop frameworks to enable statistical emulators consume data that \\ systems expose in \textit{share-data models}.\\ 
- Study and define how to apply the emulators' mathematical concept \\ in the software engineering domain.\\ 
- Exploit data exposed in DOAs to feed LLMs or transformers that can \\ inform users about system states.\\ 
- Case studies to improve our understanding of the relationships \\ between DOAs, surrogate models, and LLMs.
\end{tabular} \\ 
\hline
\begin{tabular}[c]{@{}l@{}}
Practitioners prefer centralised \\ cloud deployments based on client/\\server or microservices over \\ \textit{decentralised} architectures when \\ deploying the current ML-based \\ systems. \textbf{(RQ2)}
\end{tabular} & 
\begin{tabular}[c]{@{}l@{}}
Communities still miss tools \\ and practices around the DOA\\ style of system architecture.
\end{tabular} & 
\begin{tabular}[c]{@{}l@{}}
The lack of maturity of the DOA style negatively impacts its adoption \\ in different domains. Practitioners prefer centralised cloud deployments, \\ relying on client/server or microservices architectures. We must \\ research and develop knowledge, practices, and tools around the DOA \\ style. Alternative research paths in this direction are:\\ \\ 
- Build a knowledge base for DOA similar to the one communities have \\ developed for SOA.\\ 
- Case studies to research and quantify the limits of the DOA \\ architectural style. \\ 
- Research on the impact of the DOA style on the system's lifecycle \\ (i.e., design, development, deployment, and decommissioning).
\end{tabular} \\ 
\hline
\begin{tabular}[c]{@{}l@{}}
Systems architectures are flexible \\ at lower layers to add new sensors \\ or attend to users dynamically. \\ However, this \textit{openness} is absent \\ in the upper layers mainly because\\ systems' requirements do not \\ demand dynamic architectures at \\ these levels. \textbf{(RQ3)}
\end{tabular} & 
\begin{tabular}[c]{@{}l@{}}
The technological evolution and\\ its fast pace is exacerbating \\ systems requirements towards \\ more flexible architectures, \\ which threaten data security and \\ privacy.
\end{tabular} & 
\begin{tabular}[c]{@{}l@{}}
Future software systems will have exacerbated data requirements, \\ demanding flexible architectures where \textit{openness} plays a crucial role in \\ creating decentralised deployments and exploiting the computing \\ power of everyday devices. These flexible and open architectures \\ generate privacy and security concerns that communities must mitigate. \\ Research paths in this direction are:\\ \\ 
- Explore and exploit different approaches from the security community\\ to improve security and privacy aspects of the DOA style.\\ 
- Implement and evaluate homomorphic encryption and zero-knowledge \\ proof in the DOA context.\\ 
- Exploit decentralisation to deploy systems that store and process data \\ locally (i.e., end users' devices).\\ 
- Develop the DOA style and appropriate tools for secure deployment in \\ resource-constrained devices.
\end{tabular} \\ 
\hline
\end{tabular}
}
\end{table}

\subsection{Interplay with other areas}

\hlr{The DOA style offers a range of desirable properties for multiple software systems. The community needs additional research efforts to study the interplay of the DOA-style and computer science areas.} In this section, we mention some of the research ideas that DOA can enable in the interplay with diverse fields of study. Shared data models offer opportunities for new practices around emulating individual systems' components. Such emulation ability can lead to efficient systems' end-to-end optimisation~\cite{zeng2016joint, Aglietti2020CausalBO}. Such optimisation is particularly pertinent in cases where the system has to satisfy multiple competing requirements~\cite{Avent2020AutomaticDO}, in which multi-objective Bayesian optimisation techniques can be applicable~\cite{paleyes2022hippo}. DOA can also accelerate advances in the area of Edge Computing~\cite{cabrera2023maaco,shi2016edge,tabatabaee2022mecsurvey}, Federated Learning~\cite{bonawitz2019federated} and self-adaptive systems~\cite{cabrera2022self,gerasimou2019sefias}, all of which stand to benefit from a distributed shared data model as well as autonomous open environments. The ability to discover and inspect explicit flows of data within DOA systems has been highlighted as a key feature required for data governance~\cite{akoush2022desiderata,carata2014primer, Schwarzkopf2019PositionGC,singh2018decision}, meaning DOA can play a role in developing AI systems that comply with existing and upcoming legislatures, such as GDPR, EU AI Act or Equal Credit Opportunity Act (ECOA). Similarly, access to a complete dataflow graph means engineers can leverage causal inference techniques~\cite{10.1145/3578356.3592593,paleyes2023dataflow}.

The field of Natural Language Processing (NLP) has been experiencing rapid growth, with an increasing focus on Large Language Models (LLMs). Models such as GPT-4~\cite{OpenAI2023GPT4TR}, Llama~\cite{touvron2023llama} or Codex~\cite{Chen2021EvaluatingLL} have shown remarkable capabilities in understanding and generating human-like text or computer code. These massive models require significant computational and storage resources, making software infrastructure a key determinant of their success. Given LLMs' dependency on the quantity and quality of data, DOAs bring benefits to building efficient and robust infrastructure for them. An open research direction is to study the use of DOA principles for building LLMs and how DOA can improve their infrastructure. LLMs depend on the quality of the training data, the quality of the prompts, and the context users provide when asking questions. DOA-based systems can facilitate training and contextual data collection to fine-tune and interact with LLMs. The realisation of this vision enables promising opportunities for software generation and maintenance~\cite{robinson2024requirements} and to create novel interfaces between humans and machines to keep users in control despite the systems' complexity~\cite{cabrera2024s4}.

\subsection{Systems Monitoring and Shadow Systems}

DOA proposes that systems' components interact through data mediums. Shared data models store systems' historic states that are fully available. This data availability should facilitate systems' monitoring tasks in contrast to software design paradigms where interfaces hide data (e.g., microservices). However, in practice, we found that only one of the forty-five reviewed papers uses the shared data model for monitoring purposes. Specifically, the robotic platform proposed by \hlr{Hawes et al.~\cite{hawes2017strands} monitors historical database records of the robot status and its interactions with the environment.} This limitation emerges because communities \hlr{lack} tools for monitoring DOA-based systems. New practices and frameworks for monitoring DOA systems are needed. These practices can leverage the systems' data availability offered by the \hlg{DOA style of architecture}.

\hlr{One possible approach that fits with the key features of DOA is a network of statistical emulators (i.e., shadow systems).} A statistical emulator is a probabilistic surrogate model of a given process that allows quantifying uncertainty to inform decision-making~\cite{emukit2019}. By exposing all intermediary data interactions within the system, DOA can enable the creation and maintenance of a separate emulator for each system component. A network of such emulators acts as a shadow system, which measures the gap in behaviour between the real-world deployment and the designed software system. Such measuring includes monitoring, identifying the data stream fluctuations, and estimating the effects of potential changes. Two research efforts are required to move the idea of a shadow system beyond a mere concept. First, developing the DOA to identify the most efficient approaches and practices around the automatic fitting of surrogate models to software components. Existing work focuses on auto-tuning of system parameters~\cite{Alabed2022BoGraphSB, dalibard2017boat} and has limited scalability. \hlr{Additional use cases should implement and test shadow emulators as monitoring and explainability tools for software for exploring scalable ways of automatically building surrogate models of systems' components.} Second, innovation in the mathematical composition of individual emulators is required to build networks that can efficiently propagate uncertainty between components~\cite{Damianou2013DeepGP}. While there is prior work on uncertainty propagation in software~\cite{mishra2011uncertainty}, it is still \hlr{uncommon} to see interfaces and APIs that provide access or otherwise handle input or output uncertainty.

\subsection{DOA Concept Maturity and Tooling}

Previous sections show practitioners rely on several tools, frameworks and services while building data-driven systems. Databases, streams, message queues and file systems were all used as data communication mediums. At the same time, there is no reference that engineers could use while deciding which technology to use in their system. The adoption of DOA is now limited because the concept itself lacks maturity. Practitioners prefer centralised cloud deployments based on traditional client/server and microservices architectures. DOA can benefit from having a technical stack taxonomy so developers can have a complete list of necessary abstractions in a DOA system and a list of tools to choose for each abstraction (or their combination). Such taxonomies already exist for other areas of engineering practice, such as MLOps \footnote{AI Infrastructure Alliance project: \url{https://ai-infrastructure.org/}} or microservices~\cite{garriga2018towards}.

A question closely related to the choice of tools and frameworks is the operational maintenance of DOA systems. Over the years, the software engineering community accumulated a vast amount of knowledge on how to run SOA systems in production: what metrics to monitor, how to scale horizontally and vertically, how to mitigate and troubleshoot performance issues~\cite{lewis2008service,10.1145/2693208.2693237,beyer2016site,delac2012reliability}. A similar knowledge base about DOA systems is necessary to accumulate experience running DOA software in production. Nevertheless, researchers can study configurations of DOA setups, their performance, and critical metrics. The goal of such studies must be to determine and quantify the impact of the \hl{DOA style} on the systems' life-cycle from design to decommissioning.

\subsection{Systems Security and Privacy}

The DOA style encourages practitioners to implement open systems where autonomous entities can freely access shared data models~\cite{miao2019study,wei2021dataflow,pennekamp2019dataflow}. These data models can contain sensitive data, which naturally raises questions regarding systems' security and privacy~\cite{tsai2021privacy}. Malicious entities can access and modify systems' data and behaviour in such environments. Designers must restrict data access and systems' components to a specific set of users, considering different policies depending on the application. Decentralised deployments can mitigate security and privacy threats by storing and processing data in devices closer to end users. For example, the crowd management system proposed by Santana et al.~\cite{santana2020smartbuildings} anonymises Wi-Fi frames at the edge to protect people's identities. The authors use these frames to train a learning model for crowdsensing estimation. Despite these efforts, managing authentication, permissions, and encryption in open environments is challenging.

Different research efforts from the security community can support DOA open setups to address security and privacy challenges. Homomorphic encryption~\cite{fontaine2007survey} is an interesting direction for performing decentralised computations directly on encrypted data without providing the decryption key to the participant nodes. For example, a payment system might inquire about the transaction's validity without having access to the underlying data (e.g. bank account number). Early deployment of zero-knowledge proof~\cite{fiege1987zero} and homomorphic encryption is taking place in the industry~\cite{blum2019non}. This technology offers a solution to privacy issues in decentralised networks, but the field is still in its infancy. Algorithms developed are often computationally expensive, so further research is needed to make them more practical in resource-constrained devices. These novel research efforts must align with the prioritisation of decentralised deployment, as this DOA principle benefits privacy and security attributes by exploiting local storage and computing. In addition to these technical advances, security and privacy issues require authorities to develop novel initiatives and policy frameworks to keep up with advances in technology~\cite{montgomery2021policy}.
\section{Conclusions}
\label{sec:conclusion}

This paper surveys the extent to which, why, and how practitioners have adopted DOA to design, implement, and deploy ML-based systems in the real world. We first discuss the DOA principles based on the existing literature and explain how these can support practitioners in addressing the challenges that emerge from deploying ML models as part of larger systems. The works analysed in this survey were selected using a semi-automatic process to facilitate this study's reproducibility and evolution. We developed this process based on a well-known methodology for literature reviews in the software engineering domain. The code, raw, and preprocessed data generated in this process are available online.

Our survey shows that a few systems fully adopt the DOA principles, with most reviewed works demonstrating partial adoption. We observed that systems that follow the DOA principles are mostly data-intensive. They address requirements such as Big Data management, low-latency real-time processing, efficient resource management, security and privacy. We also show the protocols and tools these systems use for adopting the principles and distil recommendations for practitioners looking to build their data-intensive systems following DOA. We finish formulating open research challenges that can improve the community's understanding and application of DOA. We expect this work to increase community awareness of DOA and ignite interest in research, development, and adoption of this promising architectural style for real-world ML-based systems.

\begin{acks}

This research was generously funded by the Engineering and Physical Sciences Research Council (EPSRC) and the Alan Turing Institute under grant EP/V030302/1. We thank Oluwatomisin Dada, Eric Meissner, Markus Kaiser, and Peter Ochieng for the helpful and insightful discussion on this paper. 

\end{acks}

\bibliographystyle{ACM-Reference-Format}
\bibliography{ref}
\end{document}